\documentclass{aa}  

\usepackage{graphicx}
\usepackage{txfonts}
\usepackage{natbib}
  \bibpunct{(}{)}{;}{a}{}{,} 

\usepackage{xspace}   

\RequirePackage{ifpdf}
\ifpdf
  \usepackage[%
    pdftex,%
    colorlinks,%
    hyperindex,%
    plainpages=false,%
    bookmarksopen,%
    bookmarksnumbered%
  ]{hyperref}
  \definecolor{HyperColor}{rgb}{0,9,0.9,0.9}
  \definecolor{CiteColor}{rgb}{0.9,0.9,0.9}
  \hypersetup{%
    linkcolor = {HyperColor},    
    anchorcolor = {HyperColor},  
    citecolor = {CiteColor},     
    filecolor = {HyperColor},    
    menucolor = {HyperColor},    
    pagecolor = {HyperColor},    
    urlcolor = {HyperColor},     
    frenchlinks = false,         
    backref = true,
    pdftitle = {\titre},
    pdfsubject = {},
    pdfkeywords = {\MotsCles},
    pdfauthor = {A. Belu et al.},
    pdfcreator = {\LaTeX\ with package \flqq hyperref\frqq},
    pdfproducer = {pdfeTeX-0.\the\pdftexversion\pdftexrevision},
  }
\else
  \usepackage{hyperref}
\fi

\newcommand{\titre}{Primary and secondary eclipse spectroscopy with JWST: exploring the exoplanet parameter space}
\newcommand{\MotsCles}{Physical data and processes: molecular -- Techniques: spectroscopic -- Methods: analytical -- Planets and satellites: atmospheres -- Infrared: planetary systems -- Galaxy: solar neighborhood} 

\newcommand{\dgr}{\hbox{$^\circ$}} 
\newcommand{\um}  {\ensuremath{\mu\mathrm{m}}}   
\newcommand{\Msun}{\ensuremath{M_\odot}\xspace}	    

\newcommand{\Teq}{\ensuremath{T_\RM{eq}}\xspace}
\newcommand{\Tb} {\ensuremath{T_\RM{b }}\xspace}

\newcommand{\JWST}   {\textsl{JWST}\xspace}
\newcommand{\NIRSpec}{\textsl{NIRSpec}\xspace}
\newcommand{\MIRI}   {\textsl{MIRI}\xspace}
\newcommand{\LRS}    {\textsl{LSR}\xspace}
\newcommand{\HST}    {\textsl{HST}\xspace}

\newcommand{\RM}[1]{\mathrm{#1}} 

\usepackage{color}

\definecolor{NewColor}{rgb}{0., 0., 0.}
\newcommand{\New}[1]{\textcolor{NewColor}{#1}}

\usepackage{soul}
\newcommand{\Replace}[2]{ \New{#2}} 

\begin{document}

\title{\titre}
  \author{A. R. Belu      \inst{1, 2}      \and
          F. Selsis    \inst{1, 2}      \and
          J-C. Morales \inst{3}         \and
          I. Ribas     \inst{4}         \and
          C. Cossou    \inst{1, 2}      \and          
          H. Rauer     \inst{5, 6}      }
  \authorrunning{Belu et al.}
  \offprints{A. Belu}
  \institute{
    Universit\'e de Bordeaux, Observatoire Aquitain des Sciences de  
l'Univers, BP 89, F-33271 Floirac Cedex, France \email{adrian.belu@u-bordeaux1.fr} \and
    CNRS, UMR 5804, Laboratoire d'Astrophysique de Bordeaux, BP 89, F-33271 Floirac Cedex, France \and
    Institut d'Estudis Espacials de Catalunya (IEEC), Edif. Nexus, C/Gran Capit\`a 2-4, 08034 Barcelona, Spain  \and
    Institut de Ci\`encies de l'Espai (CSIC-IEEC), Campus UAB, Facultat de Ci\`encies, Torre C5, parell, 2a pl., E-08193 Bellaterra, Spain \and
    Institute of Planetary Research, DLR, 12489 Berlin, Germany \and
    TU Berlin, Zentrum f\"ur Astronomie und Astrophysik, Hardenbergstr. 36, 10623 Berlin, Germany}
  \date{Received ... / accepted ...}
  \abstract
  {During the last few years eclipse exoplanet spectroscopy havs yielded detection of H$_2$O, CH$_4$, CO$_2$ and CO in the atmosphere of hot jupiters and neptunes. In the same time, $\sim$40 likely large terrestrial planets are announced or confirmed, two of which are transiting, and another deemed habitable. Hence \New{the potential for} eclipse spectroscopy of terrestrial planets with the \textsl{James Webb Space Telescope (JWST)} has become an active field of study.}
  {We aim to extensively explore the parameter space (type of stars, planet orbital periods and types, and instruments/wavelengths) in terms of the signal-to-noise ratio (S/N) achievable on the detection of spectroscopic features with the \textsl{JWST}. We also wish to confront the information on the S/N to the likelihood of occurring targets.}  
  {We use analytic formula and model data for both the astrophysical scene and the instrument, to plot S/N contour maps, while indicating how the S/N scales with the fixed parameters. We systematically compare  stellar photon noise-only figures with  ones including detailed instrumental and zodiacal noises. Likelihood of occurring targets is based both on model and catalog star population of the solar neighborhood.}
  {The 9.6\,\um{} ozone band is detectable (S/N = 3) with \textsl{JWST}, for a warm super-earth  6.7\,pc away, using $\sim$2\% of the 5-year nominal mission time (summing observations, M4\,V and lighter host star for primary eclipses, M5\,V for secondary). If every star up to this mass limit and distance were to host a habitable planet, there should be statistically $\sim$1 eclipsing case. We also show that detection in transmission of the 2.05\,\um{} CO$_2$ feature on the $6.5\,M_\oplus$ exoplanet GJ\,1214\,b is feasible with the \textsl{Hubble Space Telescope} (\textsl{HST}). For the low and the high bound of the likely atmospheric mean molecular weight, respectively just one transit or the whole \HST yearly visibility window (107 days) is required.}
  {Investigation of systematic noises in the co-addition of 5 years worth-, tens of days separated-, hours-long observations is critical, complemented by dedicated characterisation of the instruments, currently in integration phase. The census of nearby transiting habitable planets must be complete before the beginning of science operations.}
  \keywords{\MotsCles}
  \maketitle


\section{Introduction}

Eclipsing (transit) exoplanet spectroscopy with the \textsl{Hubble Space Telescope (HST)} and \textsl{Spitzer} has enabled the detection of molecular signatures (H$_2$O, CH$_4$, CO, and CO$_2$) in the atmosphere of hot giant extrasolar planets (\citealt{tinetti-water, grillmair-addition, swainCO2, GJ436-spitzer}).  Extrasolar planets in the 1-10 Earth mass range (generally designated as ``super-earths'') have been discovered through Doppler surveys, one of which may be habitable \citep{mayor-GJ581e}. However, past decade projects for characterisation of such planets (\textsl{DARWIN} - \citealt{DARWIN_AsBio2009}, \textsl{TPF} - \citealt{allTPF}) are not technologically ready yet for implementation.
 
The recently discovered planet GJ\,1214\,b \citep{gj1214b} is the first case of close (13\,pc) transiting super-Earth, even if in this particular case the current planet density estimate points to a hydrogen-rich envelope, outside the \Replace{super-Earth}{terrestrial} regime. The prospect of extending the spectroscopy techniques above to the emerging eclipsing habitable planets, with the \textsl{James Webb Space Telescope (JWST)}, has been proposed \citep{charbonneau_deming}. The awaited performance of primary and secondary eclipse spectroscopy for habitable exoplanets is being studied (\citealt{beckwith, seager, deming_2009}, Rauer et al., submitted - RAU10 hereafter). \textsl{JWST} is scheduled for launch in 2014. 

In this paper we extensively explore the parameter space (stellar types, planet orbital period and type, and instrument/wavelength) in terms of the signal to noise ratio (S/N) achievable on the detection of spectroscopic features, for primary or secondary eclipses, with the \textsl{JWST}. Because the S/N depends on many parameters we plot contour maps, while indicating how the S/N scales with the fixed parameters. Therefore, our goal here is not to indicate the absolute performance of observations, of which we do not know yet the exact conditions, but rather identify the limits and the performance gradients over the parameter space. Also, some combinations of parameters will prove inaccessible.

\section{General target modeling}
\label{s:gen-targ-mod}

We use model stellar parameters (mass, effective temperature, surface gravity and luminosity) in the 0.1-1.4\, \Msun range \citep{baraffe}. The stars are modeled as blackbodies. We have tested on the example of a 0.3 \Msun (M3) star that the error on the S/N is below 6\% for the 5-15\,\um{} wavelength range and below 15\% in the 0.8-5\,\um{} range, when using a blackbody emission instead of a model spectrum \citep{nextgen}. Only in the 0.6-0.8\,\um{} range (which is not explored in this work), the error reaches 80\%. 

The temperature of a planet is computed as a function of its orbital distance:
 
\begin{equation}
T_\RM{eq} = \left[\frac{F_\star(a) \, (1-A)}{4 \, \sigma\, f}\right]^{^1/_4}.
\label{eq:t_eq}
\end{equation}
	 	
$\sigma$ is the Stefan-Boltzmann constant and $F_\star(a)$ is the stellar flux at the planet's location (circular orbit of radius $a$ is assumed). We assume a full redistribution of heat ($f$ = 1) for primary eclipse  observations, because we are observing the limb, which mixes flux from both high and low latitudes, as well from sunset and sunrise longitudes. For secondary eclipse, a lower redistribution factor is assumed ($f$ = 0.75), since it is the day side that is observed. The Bond albedo of the planets, $A$, is fixed at 0.2\footnote{This is the albedo the Earth would have if irradiated by a low mass star, because the emission maximum of the latter is shifted towards the infrared, where planetary molecular absorption bands are important. Also, a 0.1 difference in the albedo produces only a 3\% difference in the planet's equilibrium temperature.}  for all planet types considered.

\subsection{Planet types}
\label{s:plnt-typ}

\begin{table*}
  \caption{Parameters of planet prototypes.}
  \label{t:prm-plnt-prot}
  \centering
  \begin{tabular}{lccc}
    \hline\hline
                        & Super-Earth & ``Neptune'' &  ``Jupiter'' \\
    \hline
    Radius [Earth radii]&    2        &    3.85     &       11 \\
    Mass [Earth masses] &   10        &   18        &      317 \\
    $\mu$ (atmospheric mean molecular mass)[g mol$^{-1}]$ & 18 & 6 & 2\\
  \hline
  \end{tabular}
\end{table*}

We consider three planet prototypes, assumed to represent three large classes of planets: gas giants mainly made of H$_2$ and He, icy giants (Neptune-like), and large terrestrial planets. A planet prototype is defined by its mass, radius, and, for primary transit observations, its atmospheric mean molecular weight $\mu$.

\paragraph{``Jupiters''.} As presented in the introduction, both primary and secondary eclipse spectroscopy for hot jupiters is being achieved today. Future instruments will give access to higher spectral resolutions, and cooler planets, so accordingly we consider a Jupiter-mass planet for the present study, with $\mu$ = 2\,g\, mol$^{-1}$ (mainly H$_2$). 

\paragraph{``Super-earths''.} Terrestrial planets can be indicatively defined as having an upper limit on their mass of 10 Earth masses, although in particular cases planets with a slightly lower mass can accrete a massive gas envelope (\citealt{rafikov} for modeling, also probably the case for GJ\,1214\,b). Therefore the prototype of a terrestrial planet considered in this study is a 10 Earth mass-, 2 Earth radii-planet (hereafter ``super-earth''). 

Our super-Earth prototype is considered habitable when found within the limits of the circumstellar habitable zone, as defined by \citet{sels_gl581}. For all terrestrial planets, we consider an optimistic $\mu = 18$\,g\,mol$^{-1}$ (water vapor dominated atmosphere of a planet in the inner portion of the habitable zone), instead of Earth's 28 (N$_2$ dominated).

\paragraph{``Neptunes''.} Neptune-mass planets represent an intermediate between the habitable case and the Jupiter-mass case. For this prototype we consider $\mu$ = 6\,g\,mol$^{-1}$ (H$_2$, He and 10\% of heavier elements). 

The mean molecular weight we use is actually higher than that of Neptune and Uranus, which is close to 2\,g\,mol$^{-1}$. The reason is that atmospheric escape is likely to deplete the amount of hydrogen on hot neptunes. The escape parameter for a given species $i$ is $X_i = R_\RM{p}/H_i$ where $R_\RM{p}$ is the radius of the planet and $H_i$ is the individual scale height of the species $i$, calculated at the exospheric temperature. Escape becomes important for $X < 15$, while the gas is tightly bounded to the planet for $X\,>\,30$. For Neptune and Uranus, the exospheric temperature is 700-800\,K and $X_\RM{H}$ is in the range 35-45. For warmer planets, the exospheric temperature of a Neptune-like planet can be much higher, as it roughly scales linearly with the stellar flux, until significant thermal ionization occurs, so up to a few thousand K \citep{lammer}. Exospheric temperatures above 2,000\,K can safely be assumed for warm and hot Neptune-type planets. This would result in values of $X_\RM{H}$ below 15, and thus to a rapid escape of hydrogen. Therefore, for the hottest Neptune-type planets able to keep an atmosphere, the remaining atmosphere should be enriched in heavy elements or even consist mainly of heavy molecules (like N$_2$, CO$_2$, H$_2$O). Therefore, our choice of 6\,g\,mol$^{-1}$ is an average situation where only part of the hydrogen is left and should overestimate (respectively underestimate) $\mu$ for planets with a long (respectively short) orbital period.
\\

\noindent Table~\ref{t:prm-plnt-prot} summarizes the parameters of our planet prototypes. Note that the $\sim$40 transiting exo-jupiters detected until now have a wide variety of densities, both lower and higher than that of Jupiter. Since we need to limit the number of parameters, we chose the Jupiter parameters as a middle case for the ``Jupiter'' prototype.

\subsection{Spectral signatures considered and their modeling}
\label{s:spec-sigs}

S/N calculations can be based on the spectral features found in Solar  
System planets but this approach covers only a negligible fraction of  
the parameter space. They can also be based on synthetic spectra  
computed for specific atmospheric composition. In such case, the  
structure and composition of the atmosphere has to be modeled self-consistently by coupling radiative transfer, (photo)chemistry and  
dynamics. For the atmosphere of giant gaseous planets, elemental  
composition should not depart too dramatically from the stellar  
composition, although selective enrichments and depletion are expected to occur due to the separation of condensed and gaseous phases, or to  
gravitational escape,  in the protoplanetary disk and in the planet.  
For these planets, it is thus conceivable to produce grids of  
spectra covering a limited number of parameters, as it is done for  
stars. But even in this case, producing such grids would imply some  
drastic simplifications (1D instead of 3D, equilibrium chemistry  
composition instead of kinetics and photochemistry, simple cloud  
models, decoupling of radiative, dynamical and chemical processes) and would suffer from the incompleteness of  
the required physical/chemical data (spectroscopic data, kinetic  
rates).

For low-mass rocky and icy planets, the situation is extraordinary  
more complex.  Their spectral properties are determined by an  
atmosphere initially accreted as volatiles trapped in solids, or ices, 
of non-solar composition. This volatile content represents a small  
fraction of the total planetary mass and is fractionated between the  
interior (crust, mantle), the surface oceans and/or ice sheets, the  
atmosphere and outer space through gravitational escape (induced by  
impacts, exospheric heating and non-thermal processes). The  
composition of the atmosphere of a terrestrial planet at a given stage of its evolution is then controlled by geochemical exchanges between  
these different reservoirs, tectonics, atmospheric escape,  
photochemistry, and biology if present.
 
Therefore, the expected diversity of exoplanet atmospheres, and  
terrestrial planets in particular, covers a wide parameter space, and  
our current understanding of the origin and evolution of planetary  
atmospheres provides very few constraints to guide us in this  
exploration. Although the use of detailed atmosphere models and  
synthetic spectra is essential, in particular to interpret spectral  
observations, it is equally important to allow ourselves to explore a  
much broader parameter space than the one covered today by 
self-consistent models. 

This is why we chose to base this study on a different, ``model-less''
approach, which is complementary to the use of detailed atmosphere  
models, which remains necessary to refine the actual S/N for a  
specific close-up in the parameter space (for instance RAU10, based on self-consistent habitable planet atmosphere models).

Because of the reasons above, we chose here to examine the S/N of \emph{individual} features of species, freeing us from any a priori on the atmospheric composition and structure. Moreover, what interests us here is not the absolute planetary signal flux, but the the \emph{detection} of a spectral feature. 

Therefore, we model the detection by estimating the difference of the planetary flux between two appropriately chosen binned channels, one measuring the continuum, and the other the flux in the absorption band of the feature. Of course, when a given (photo)spectroscopical observation comprising up to tens of channels will be fitted with synthetic spectra, the S/N on the detection of species will be much higher. With this definition, an S/N of 3 is a safe $3\,\sigma$ detection (also see Section~\ref{s:spec-feat-sig} below).

In general, we chose to compute the S/N for a fiducial signature defined by a given spectral resolution, and a contrast necessary for its detection. The way the signature contrast (or depth) is defined is described in the next section. 

However, we also wish to particularly emphasize the case of the habitable super-earths. As such, we consider some of the strongest infrared signatures of species present in the terrestrial atmosphere:

\paragraph{CO$_2$ feature at 4.3\,\um{}.} We measure this feature relative to a region redward of 4\,\um{}. Therefore, we consider in the calculation a mean working wavelength of 4\,\um. (and an effective width of 0.4\,\um, so $R\,=\,10$).

\paragraph{CO$_2$ feature at 15\,\um{}.} Since this a filter observation for \JWST, the modeled width of the feature will be specified in the appropriate section below (Section~\ref{s:miri-first}).

\paragraph{O$_3$ feature at 9.6\,\um{}. } The considered width is 0.5\,\um{} (so $R\,=\,20$).

\subsection{Types of transits}

Depending of the type of transit, we use several assumptions to compute the planetary spectral feature depth.

\subsubsection{Primary transit.}

\begin{figure}
  \centering
  \includegraphics[width=\columnwidth]{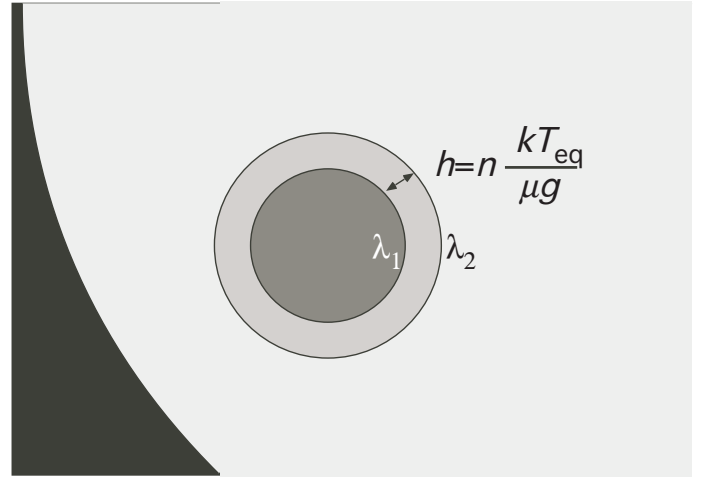} 
  \caption{\label{f:primary}Depth of a planetary spectral feature in transmission  (opacity height difference). $\lambda_2$ is the feature's central wavelength, and $\lambda_1$ is the reference channel wavelength (continuum). $k$ is the Boltzmann constant, $T_\RM{eq}$ the equilibrium temperature of the atmosphere, $\mu$ its mean molecular weight, and $g$ the surface gravity. $n$ is the relative strength (usually $n=3$, see text).}
\end{figure}

We use the same formulas as \citet{beckwith} for the planetary spectral feature photon count (we consider the additional background and instrumental noises as indicated in Eq.~\ref{eq:s2n}). We chose the difference in atmospheric opacity height between the in- and out-of-band channels to be $n\,=\,3$ \footnote{Change of 1 order of magnitude in atmospheric pressure} atmospheric scale heights $H = k\,T_{\RM{eq}} /\mu{}\,g$ (Figure~\ref{f:primary}), $k$ being the Boltzmann constant.  Consequently, the S/N scales with $^1/_\mu$ and $n$. This value has been observed for hot jupiters between adjacent spectral bins (even though larger differences in the apparent radius have been measured over entire spectra, see previous discussion on S/N definition on this page). Another way of seeing our modeling is as an achievable ``resolution in amplitude''. A ``$n\,=\,3$ sampling'' should actually enable to detect opacity-radius variations of the planet over extended wavelength ranges (i.e. spectra) with a ``bit depth'' that could be handy if disentanglement of the signatures of multiple species is required). $n\,=\,3$ is also a high value for the Earth case, where greatest opacity height difference is $4\,H$ for the 15\,\um{} CO$_2$ and the 9.6\,\um{} O$_3$ bands, and $5\,H$ for the 4.3\,\um{} CO$_2$ band \citep{lisa-transits}.

\subsubsection{Secondary transit}

Secondary eclipses allow to investigate both thermal emission and star reflection from the planet. We wish to have an idea of the weight of each phenomenon in the planetary flux, although they are undistinguishable in an observation. 

\begin{figure*}
  \centering
 \includegraphics[width=\textwidth]{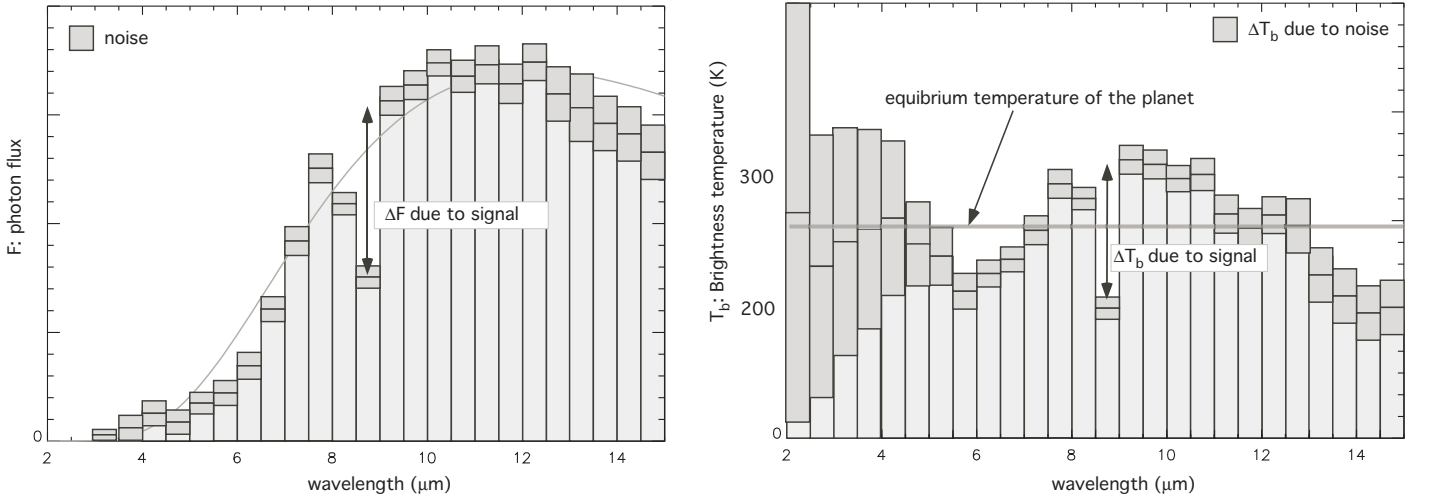} 
  \caption{\label{f:bright-temp}Depth of a spectral feature in emission (see text for discussion).}
\end{figure*}

\paragraph{Emission.} The radius of a transiting planet is known from the primary transit at short  wavelengths, where the thermal emission from the night side of the planet can be safely neglected. The emitted spectrum measured  at the secondary eclipse can be converted into a brightness  temperature, $T_\RM{b}(\lambda)$. The depth of spectral features in the emission  spectrum depends on the vertical temperature profile (a isothermal atmosphere emits a featureless blackbody spectrum whatever its composition). At a given wavelength, the signal comes from the altitude $z_\lambda$ corresponding to an optical depth $\tau_ 
\lambda \approx 1$, and the associated brightness temperature is the physical temperature at this level: $T_\RM{b}(\lambda)=T(z_\lambda)$. The altitude $z_\lambda$ is a function of the abundance profiles of molecular absorbers. For a given wavelength domain, the amplitude of the temperature variations within the altitude range spanned by $z_\lambda$ gives the upper limit on the depth of the observable features in the brightness temperature spectrum. Emission features can however be even stronger for a non-LTE atmosphere, where for instance fluorescence occurs, like it might be the case for HD\,189733\,b \citep{ground-swain}.

The spectral signal we want to detect within the noise can thus be expressed as a ``resolution in amplitude'' (precision) to be achieved in measuring the brightness temperature (at the spectral resolution of the considered feature). We therefore consider the depth of the spectral feature in emission to be the difference in the measured planetary thermal emission between the in- and out- of feature channels (Figure~\ref{f:bright-temp}).

For the fiducial signature, in order not to be compelled to  
any assumption on the detailed composition and vertical structure of  
the atmospheres, we chose to define the aimed resolution as a fraction $\alpha\,=\,20\%$ of the equilibrium temperature computed for the planet.  
Compared to the Earth case ($T_\RM{eq}\,=255\,\RM{K}$), $\alpha\,=\,20\%$ corresponds  
to about the highest temperature contrast ($\Delta{} T_\RM{b}\,=\, 50\,\RM{K}$)  
that can be observed at low resolution on a disk-averaged spectrum. For this observation, $T_\RM{b}$ can reach near-surface temperatures of about 270~K in  
the 10-11\,\um{} atmospheric window, and 220~K in the 15\,\um{} CO$_2$  
band probing the lower stratosphere (see for instance \citealt{obs_terre_mgs}). Slightly higher relative contrasts of brightness temperature have been modeled for hot and habitable Super Earths (RAU10) and significantly higher relative contrasts have actually been observed on hot exoplanets \citep{madhusudhan-seager}. This value of $0.2\,T_\RM{eq}$ for the brightness temperature resolution appears thus as a reasonable goal to study exoplanets in general, and  
possibly the minimum required precision to search for atmospheric  
signatures habitable planets.

We chose \Teq to be the midpoint of the brightness temperature variation, and not the upper bound (continuum temperature). Indeed, depending on the wavelength, atmospheric or surface temperatures probed by the observations can be higher or lower than \Teq, which is a mean value. For instance, the disk averaged continuum emission of the Earth in the 8 and 12\,\um{} windows has a \Tb significantly higher than \Teq (due to the greenhouse effect), except within the 9.6\,\um{} O$_3$ band, where \Tb is lower than \Teq.

The 4.3\,\um{} CO$_2$ feature is modeled in emission as having  $\Delta{}T_\RM{b} = 100\,\RM{K}$ \citep[Fig. 8.9]{thesis_paillet}. The 15\,\um{} CO$_2$ and the O$_3$ features are modeled with $\Delta{}T_\RM{b}$ respectively 60 and 30\,K. However, around low-mass stars, the 15\,\um{} CO$_2$ band can be as deep as 100\,K (RAU10). Also, O$_3$ is the main absorbent responsible for the stratosphere temperature inversion;  some temperature profiles and O$_3$ mixing ratios may produce a signature stronger or weaker than 30\,K. It must be noted however (valid also for the primary transit case), that a habitable super-Earth around a low-mass star may require a dense CO$_2$ atmosphere in order to keep an atmosphere at all \citep{joshi,scalo}. This CO$_2$ may swamp the O$_3$ signal \citep{selsis_2002}.

\paragraph{Reflection.} For reflected light spectroscopy at the secondary eclipse, we define our signal as a 50\% difference of the specific (i.e. $\lambda$-dependent) planetary albedo, between the in- and out-of-feature channels. This is considered at 1\,\um{}, and for a feature width of 0.1\,\um{} ($R$ = 10).

\subsubsection{Invariance of the primary transit photon count.}

Interestingly enough, we note that the fundamental physical information for primary transit transmission spectroscopy, that is, the number of stellar photons traversing an atmospheric scale height, for a single transit, is a quantity independent from the planet's period $P$. 

Since, $F_\star(a) = \sigma{}T_\star^4 \, \left(R_\star/a\right)^2$, with $R_\star$  being the stellar radius and $T_\star$  the stellar temperature, $F_\star \propto P^{-4/3}$, where $P$ is the orbital period of the planet, therefore $T_\RM{eq} \propto P^{-1/3}$. For primary transits, the solid angle of the opaque annulus corresponding to an absorption spectroscopic feature, over a given bandwidth, is $\Delta\Omega = 2 R_\RM{p} \, 3H/d^2$. So $\Delta\Omega \propto P^{-4/3}$. The planetary spectral signature flux is $F_\RM{p}(\lambda) = \Delta\Omega \, B(\lambda,T_\star)$, where $\lambda$  is the wavelength and $B(\,,\,)$ is the Planck function. So $F_\RM{p}(\lambda)\propto P^{-1/3}$. The photon count for a planetary transit is $n_\RM{p} \propto F_\RM{p}(\lambda) \, \tau$ with $\tau$  the transit duration. $\tau = P R_\star \, / \, \pi{}a$, so $\tau \propto P^{1/3}$. So $n_\RM{p}$ is independent of $P$ (in the frame of the approximations above).

\section{Signal of the spectroscopic feature and noises}
\label{s:spec-feat-sig}

In a simple model of exoplanet eclipse observation, the planetary flux is the difference between the estimates of the in-transit flux and the out-of-transit one (both assumed constant with time). Assuming absence of correlation between the two (stellar photon noise dominated), the variance of the planetary flux is therefore the sum of the variances of the two estimates. Current observations of a single eclipse already have calibration precisions of the same order as the stellar photon noise, whether they cover the out-in-out sequence with multiple telescope pointings  \citep{swain_ch4} or a single one \citep{grillmair2007}. If we acquire photon counts for the transit and outside of the transit over the same maximum available time period (i.e the transit duration), these two photon counts can be considered as having same variance. Increasing the out-of-transit integration time reduces the variance of this term, in comparison with the in-transit one, if the only variability source is the stellar photon noise. Stellar oscillations however contribute to the variance of the constant star flux estimate.  This is why model fitting is normally used to estimate the planetary signal in transit spectroscopy observations, and should be considered in a future iteration of this work. Post-detection methods for stellar variability filtering \citep{alapini} could be particularly well suited for massively co-added transits, around the active, lowest mass M dwarves (see below).

Also, we have seen previously that we consider the difference of the planetary flux between the in-feature- and the out-of-feature channels. Again, the variance on this estimate is the sum of the variances of the two terms (supposing uncorrelated noise between channels, such as photon noise). The in-feature binned channel width, $\Delta\lambda$, is constrained by the width of the feature (for instance 0.5\,\um{} for the O$_3$ band at 9.6\,\um{}). The out-of-feature binned channel (possibly divided in two, flanking the feature), is chosen depending on the profile of the spectrum (optimistically assuming that a clean continuum can be defined).

Following these considerations we chose a simplified model where the out-of-transit observation time is equal to the duration of the transit. We also consider equal out-of-feature and in-feature binned channel widths. We thus compute a signal-to-noise-ratio on the spectral feature \emph{detection}:

\begin{equation}
S/N = \frac{\RM{planetary~spectral~feature~photon~count~for~1~transit}}{\sqrt{4\times\left(\sigma_{n_\RM{star}}^2+\sigma_{n_\RM{zodi}}^2+\sigma_{n_\RM{thermal}}^2+\sigma_{n_{RON}}^2+\sigma_{n_\RM{dark~current}}^2\right)}} \, ,
\label{eq:s2n}
\end{equation} 
where $n$ are the photon counts for each subscript source, for the duration of a transit (``zodi'' stands for the zodiacal light contribution, ``RON'' for readout noise). Exo-zodiacal light was not modeled. First, the contribution to the noise of an exozodiacal cloud (viewed in its entirety), similar to the solar one, is negligible when compared to the stellar photon noise. For systems with very high dust levels, its brightness variability over the time scale of the observation has no foreseeable source. Second, little is known today about the statistics of exozodiacal dust densities around nearby stars, which however an ongoing effort \citep{KIN-2009}.

The \textsl{JWST} features a primary mirror of $D\,=\,6.5\,\RM{m}$ diameter and a throughput before instrument of 0.88 \citep[hereafter DEM09]{deming_2009}. It will be equipped with several instruments potentially enabling the molecular eclipse spectroscopy we are considering\footnote{Deming et al. Exoplanet Task Force White Paper, 2008, table at end.}. Furthermore, each instrument has different observation modes (filter photometry, as well as low and intermediate resolution spectroscopy), that can be used for exoplanet spectral characterisation.

\subsection{NIRSpec}

In the near-infrared (NIR), we consider the \NIRSpec instrument \citep{rauscher}. Its performance for primary transit spectroscopy in the $R\,=\,1,000$ mode, for the detection of water and CO$_2$ has already been studied (DEM09). We therefore focus here on the $R\,=\,100$ mode (0.6-5\,\um), which could potentially yield a higher throughput and less readout noise, but at the expense of more saturation. This mode is therefore better suited for the faintest target stars, around which spectral characterization of super-earths will be most efficient. We actually show below that the saturation will not be a limiting factor, considering the number of target stars of a given type within a given distance from the Sun. 

The overall throughput (including quantum efficiency) and the resolution function were provided by P. Ferruit (CRAL, Lyon, France). Since exoplanet transit observations will be done in a pseudo slitless mode (custom 1.6\arcsec$^2$ opening), we scale the provided resolution function (which was computed for the 0.2\arcsec{} slit) with the point-spread-function (PSF) $\lambda$-dependent size. For wavelengths under 1\,\um{} we use the PSF size at this upper bound, to account for distortions that become non negligible below this value. The resulting resolution curve serves to compute the number of pixels in the spectral feature channel, hence the readout noise. The pixel scale is 0.1\arcsec, the readout noise is 10\,e$^-$/pix\,rms, and the well capacity is 60,000\,e$^-$. Since the PSF is undersampled at the shortest wavelengths, we suppose that the spectrum axis is centered between two rows of pixels. A defocus mechanism could however be present to mitigate the undersampling problem.

In order to compute the brightest pixel, we use the resolution function at the shortest wavelength (maximum star emission), by assuming that there is no diffraction in the dispersion direction. This is an accurate model in terms of energy distribution over short dispersions which are part of a larger spectrum (the resolution function is computed for a 2.2 pixel size of the resolution element). In the spatial direction, we use a simple triangle model of the center of our dispersion-anamorphosed PSF (83\% of the total energy). The maximum number of electrons in the brightest pixel is used to compute the readout rate. 

To determine the read time, we compute the length of the \NIRSpec $R\,=\,100$ spectrum from the resolution function (356.2\,pixels), rounded to the upper power of 2 (512). The width (in the spatial direction) on the detector is 2 $\times \lambda_\RM{max}/D$ ($\lambda_\RM{max}\,=\,5$\,\um{} here), rounded in pixels to the closest upper power of 2 (8 pixels). We assume the read mode is MULTIACCUM-2$\times$1 \citep{rauscher}, meaning that we only have to account for the reset frame time (the detector is read non-destructively up-the-ramp). We reduce the effective photon collection time over the transit by this amount (the remaining fraction is called \textit{duty cycle}).

\begin{figure}
  \centering  \includegraphics[width=0.7\columnwidth]{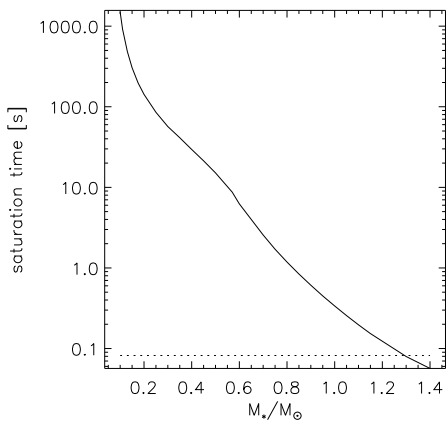}
  \caption{\label{f:sat-nirspec}Brightest pixel saturation time for the full $R\,=\,100$ \NIRSpec mode, as function of star mass (star at 10 pc). Dotted level: spectrum window reset time.}
\end{figure}

Figure~\ref{f:sat-nirspec} shows the brightest pixel saturation time as a function of stellar type, and the full spectrum window reset time level, assuming each pixel requires~10\,$\mu$s to reset. Smaller detector windows can be defined (possibly alternately observed), depending on the sought spectral signatures. However, we have not modeled such observations, so we limit our \NIRSpec plots at the bound defined by the intersection of the reset time and the saturation time: 1.4\,\Msun at 10\,pc (four F4 stars only: Sirius A, Altair, Formalhaut, Vega and Procyon A). The limit is 1.05\,\Msun at 5 pc (three stars only: $\alpha$  Cen, Sirius A and Procyon A).

Finally, the PSF being undersampled at the shortest wavelengths, the spectrum's jitter over the pixel grid (7 mas rms, DEM09) is likely to constitute the major noise source in the NIR; however, this modeling is beyond the scope of this work. For instance, mapping all detector pixels in the dispersed custom window, before flight, should enable processing techniques that reduce this noise contribution. 

\subsection{MIRI}

The mid-infrared (MIR) range is covered by the \textsl{Mid-InfraRed Instrument} (\textsl{MIRI}, \citealt{MIRI_Wright}). Its filter photometry mode performance has been already studied for the detection of the CO$_2$ band at 15\,\um{}, for secondary transit spectroscopy (DEM09). However, the current filter set is not optimal for the detection of the 9.6\,\um{} O$_3$ band (the \textsl{IM\_3} imaging filter covers 9-11\,\um{}, four times the width of the feature). We therefore consider here the 5-11\,\um, low resolution ($R = 100$) spectrometer mode (\textsl{LRS}). The optical transmittance is assumed to be 0.4 for both modes. Both modes use the same detector with a pixel scale of 0.11\arcsec, a readout noise of~19\,e$^-$/pix rms, and a well capacity of 10$^5$ e$^-$. Again, we do not consider the effect of the aforementioned instrumental jitter on \MIRI (PSF undersampled shortward of 7\,\um), although it was checked in filter mode on one simulation that its effect is negligible compared to other limitations (Cavarroc et al., forthcoming).

We assume a total spectrum length of 194 pixels in the dispersion direction (wavelength pixel registering provided by S. Ronayette, CEA). We make the assumption that the read mode is similar to that of \NIRSpec (above).

The extension of the feature on the spectrum is computed by a simple proportionality between the width of the spectral feature  and the total length of the spectrum. No saturation occurs for \MIRI at 10 pc over the considered 0.1-1.4\,\Msun range of stars.
\\
\\
For both instruments, a 0.03\,e$^-$\,s$^{-1}$\, pix$^{-1}$ dark current noise is considered. Uniform background noise sources are calculated using the pixel scale of the detectors. For the instrument's thermal emission we use a temperature of 45\,K and a 0.15 global emissivity.

\subsection{Zodiacal light}

We use an implementation by R. den Hartog of a parametric model by O. Lay of the Kelsall local zodiacal cloud model \citep{darwinsimref}. The most pessimistic ecliptic latitude ($\beta$ = 0) is used, but in the anti-solar direction, which represents a good average of the all-sky distribution.

\section{Results}

We now present the computed S/N for the different types of transits, for different types of planets, and for different wavelength ranges. For the smallest planets, achieving a significant S/N will require cumulating data from multiple transits, in which case the S/N scales with the square root of the number of transits, provided that (instrumental) noises are not correlated between the successive transit observations. Hence, the detection limit is ultimately the mission life-time (5 years), which limits the S/N that can be achieved on longer period planets. It must be emphasized that such an hypothetical observation, while having a total observation time only a magnitude over the longest exposures made until now \citep{HUDF}, suffers from the risks inherent of being distributed over a 100 times larger duration. In other words, if \JWST becomes inoperable after 2.5 years of operations, the 1\% of the mission time dedicated to acquiring data on the planet would be lost, since, the data would yield an insufficient S/N.

It must be noted that the yearly target visibility for \JWST reaches 100\% only for targets with ecliptic latitude higher than 85\dgr. The visibility is lower than 100 days per year for latitudes up to 45\dgr \footnote{\href{http://www.stsci.edu/jwst/overview/design/field_of_regard.html}{http://www.stsci.edu/jwst/overview/design/field\_of\_regard.html}}. We find the  yearly mean sky visibility to be 149 days. For all following super-earths plots, we therefore multiply the number of transits occurring over the 5-year mission time by the corresponding fraction, and obtain the effective number of observable transits. For all planet types, we also ponder for the unknown impact parameter of the transit, by further multiplying the transit duration (equatorial) by $\pi/4$ ($\sim$0.79).

Unless mentioned otherwise, all our examples here are calculated for a system at a distance $d$ = 10\,pc, the signal to noise scaling linearly with the inverse of the distance. Jupiter-mass planets are studied at 50\,pc (see Section~\ref{s:stats}).

\subsection{Primary transit}

The main interest of primary transit spectroscopy is that it yields a spectrum even in the case of an atmosphere which is isothermal, or  which has a low temperature gradient (this would not be case if this atmosphere where observed in emission, during a secondary transit). \New{Also, the primary transit signal is proportional to $T_\RM{eq}$, whereas this is true in emission only in the Rayleigh-Jeans regime. As a consequence, some ($\lambda$, $T_\RM{eq}$) couples (cold planets at short wavelengths) may have an undetectable emission flux, but could be characterized in primary transit (see below).}

\subsubsection{NIRSpec}

\begin{figure*}
  \centering
   
  \includegraphics[width=0.49\textwidth]{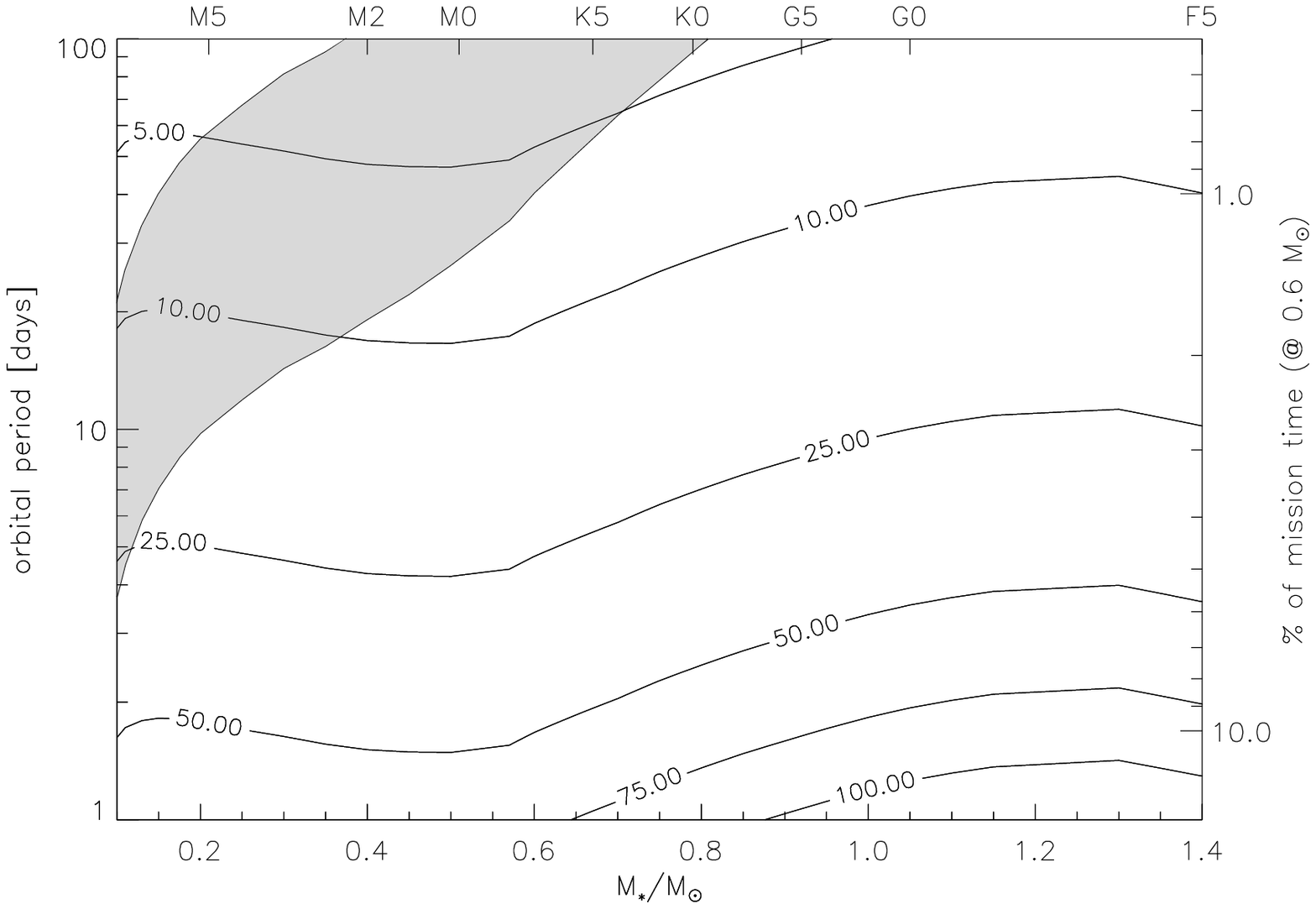}  \includegraphics[width=0.49\textwidth]{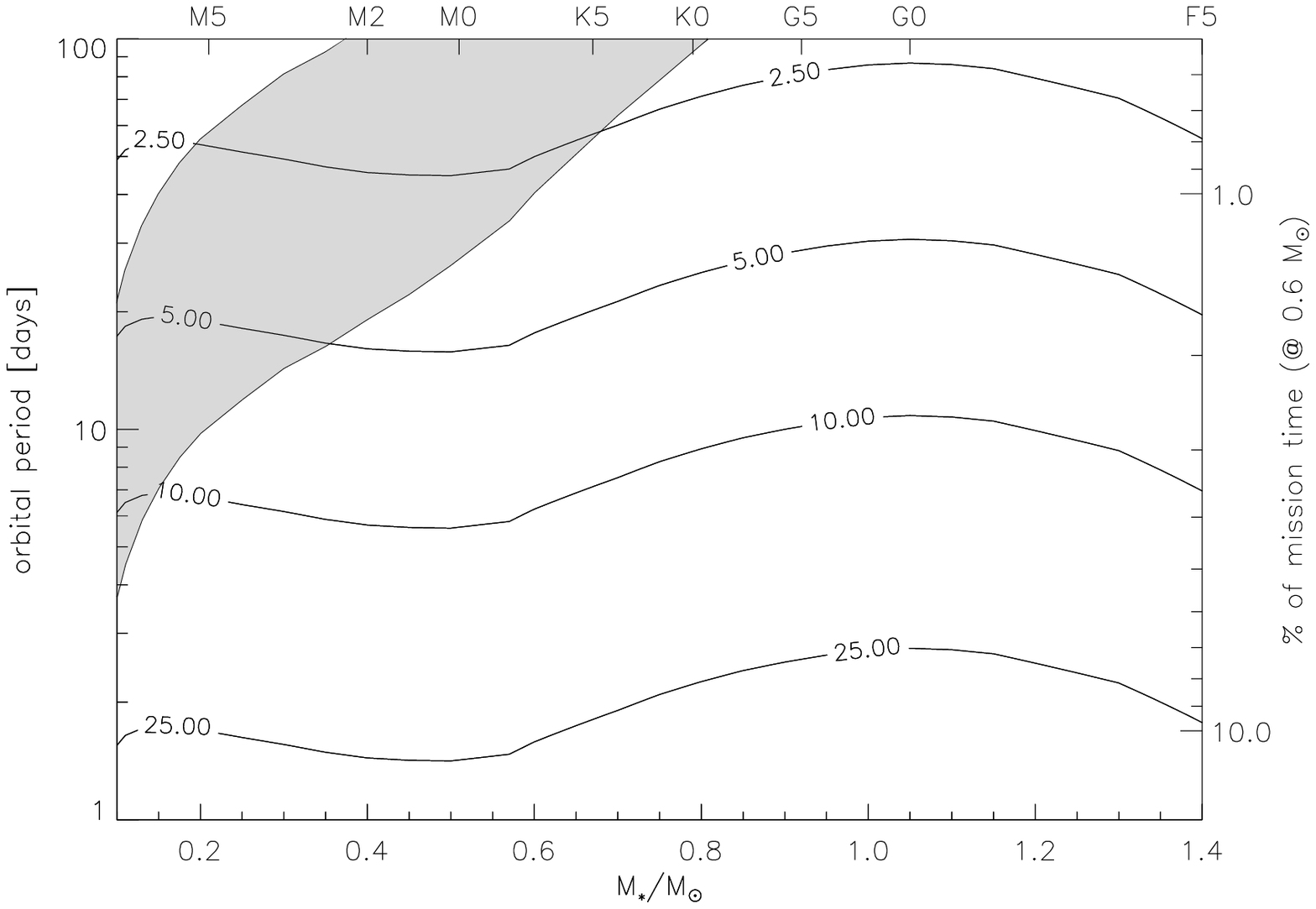}
  
  \caption{\label{f:4.3-CO2-trans}S/N on the detection of the 4.3\,\um{} CO$_2$ signature in a super-Earth at 10\,pc, for observations of all the primary transits available on average over the 5 year mission time. \textbf{For this and all following S/N plots}: \textit{left} plot is with stellar noise only and \textit{right} plot is with instrumental (here \NIRSpec) and zodiacal noises. \textbf{For this plot and following cases of observations over the whole mission time (super-earths)}: a)\,the habitable zone is plotted in gray, b)\, the fraction of the mission time, accounting for 2 transit durations ("1/2 out + in + 1/2 out-of-transit"), plus the fixed 65\,min \JWST slew time, is on the right axis.}

  \includegraphics[width=0.49\textwidth]{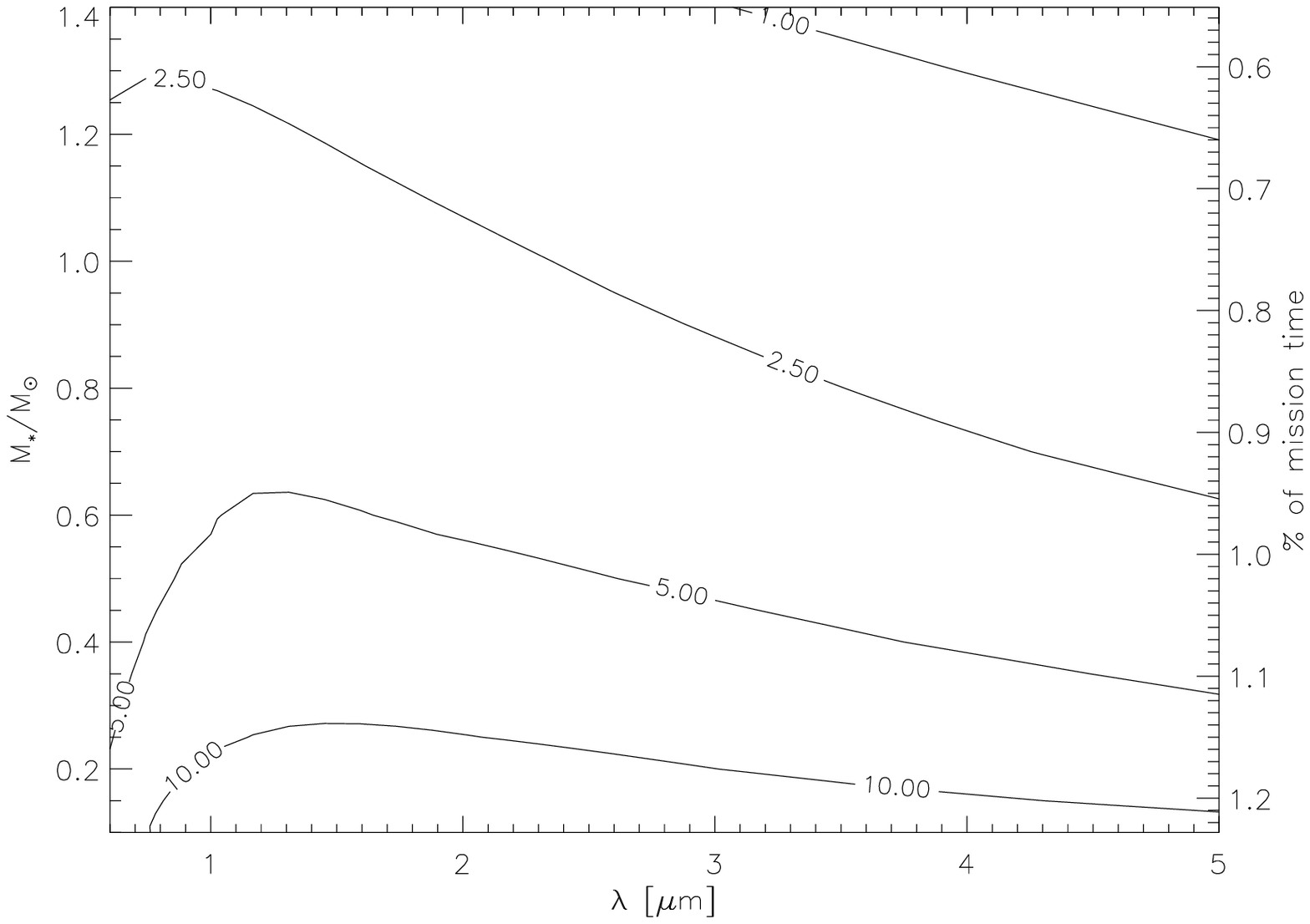}   
  \includegraphics[width=0.49\textwidth]{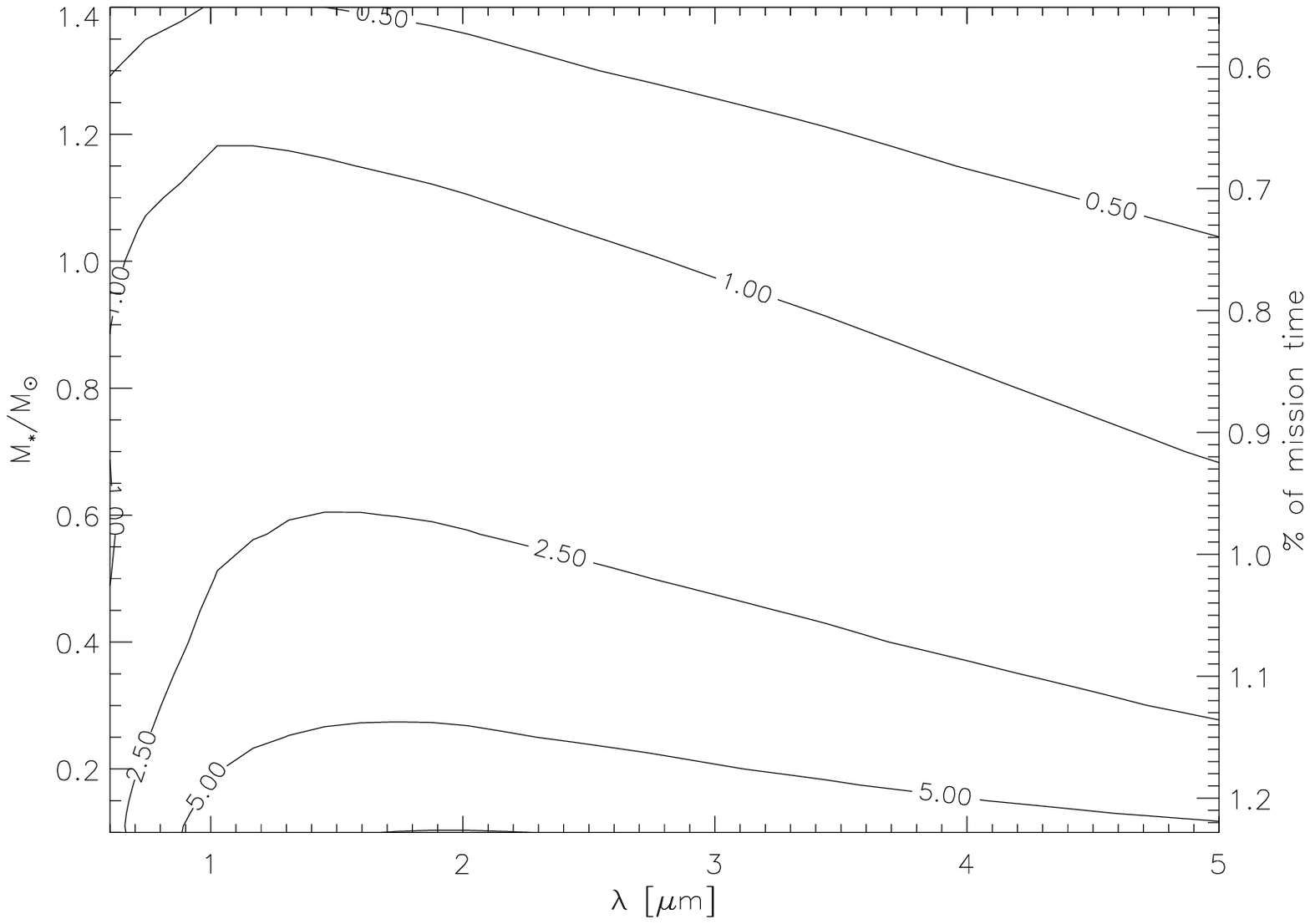}
  
  \caption{\label{f:se-hz-1}Same as Figure~\ref{f:4.3-CO2-trans} but in the (wavelength - star mass) parameter space. The planet's distance to the star is such that it receives the same amount of energy as Earth (1\,AU from the Sun). The width of the fiducial spectral feature used in the computation corresponds at each wavelength to a fixed resolution ($R=20$).}

\end{figure*}

\noindent Figure~\ref{f:4.3-CO2-trans} shows the S/N for the detection of the 4.3\,\um{} CO$_2$ band as modeled in Section~\ref{s:spec-sigs}, for a super-Earth (this figure serves as a template for following similar plots; refer to its caption for conventions). All primary transits available on average (see beginning of current section) are cumulated over the 5 year mission time, with stellar noise only (left) and the instrumental and zodiacal noises (right). We have checked on several examples that our figures with stellar noise only compare well with RAU10. The main noise influence here is the readout-noise. CO$_2$ can be detected on super-earths in transmission. The distance to the star being fixed in this type of plot, the decrease of S/N towards  brighter (heavier) stars when considering instrumental noises is due to the increasing weight of the reset time in the duration of each exposure, hence the reduction of duty cycle. We limit our study to the main sequence dwarves M through F. While the latter, more extended stars should provide an increased duration for collecting photons (transit duration), the duty cycle displays a maximum at 0.6\,\Msun. The position of this maximum is 0.5\,\Msun (M0) at 5\,pc.

In order to explore the sensitivity of the instrument at various wavelengths (for other potential spectral signatures), we consider the fiducial spectral signature ($n = 3$ scale heights $H$, and of constant spectral resolution of 20 over the wavelength range - Figure~\ref{f:se-hz-1}). The S/N scales linearly with the inverse of the square root of the resolution. The dominating noise source is the readout noise, set by the saturation time of the brightest pixel of the spectrum. The variation with stellar type of the wavelength of maximum emission is clearly visible. The abrupt reduction of \NIRSpec throughput towards the shorter wavelengths is also visible when comparing the star-only-noise plot (left) with the full modeled noise one (right).

\begin{figure*}
  \centering  
  \includegraphics[width=0.49\textwidth]{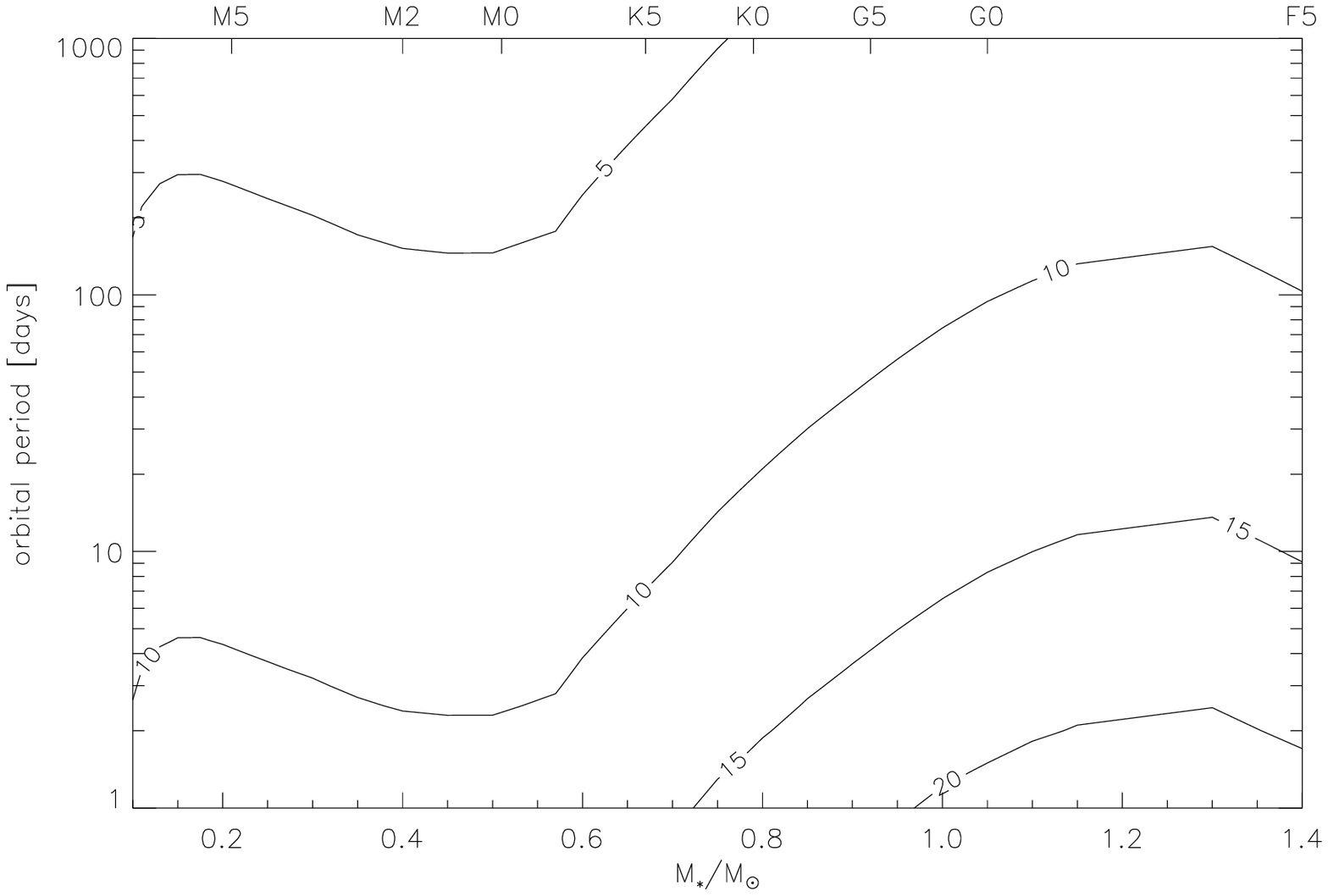}
  \includegraphics[width=0.49\textwidth]{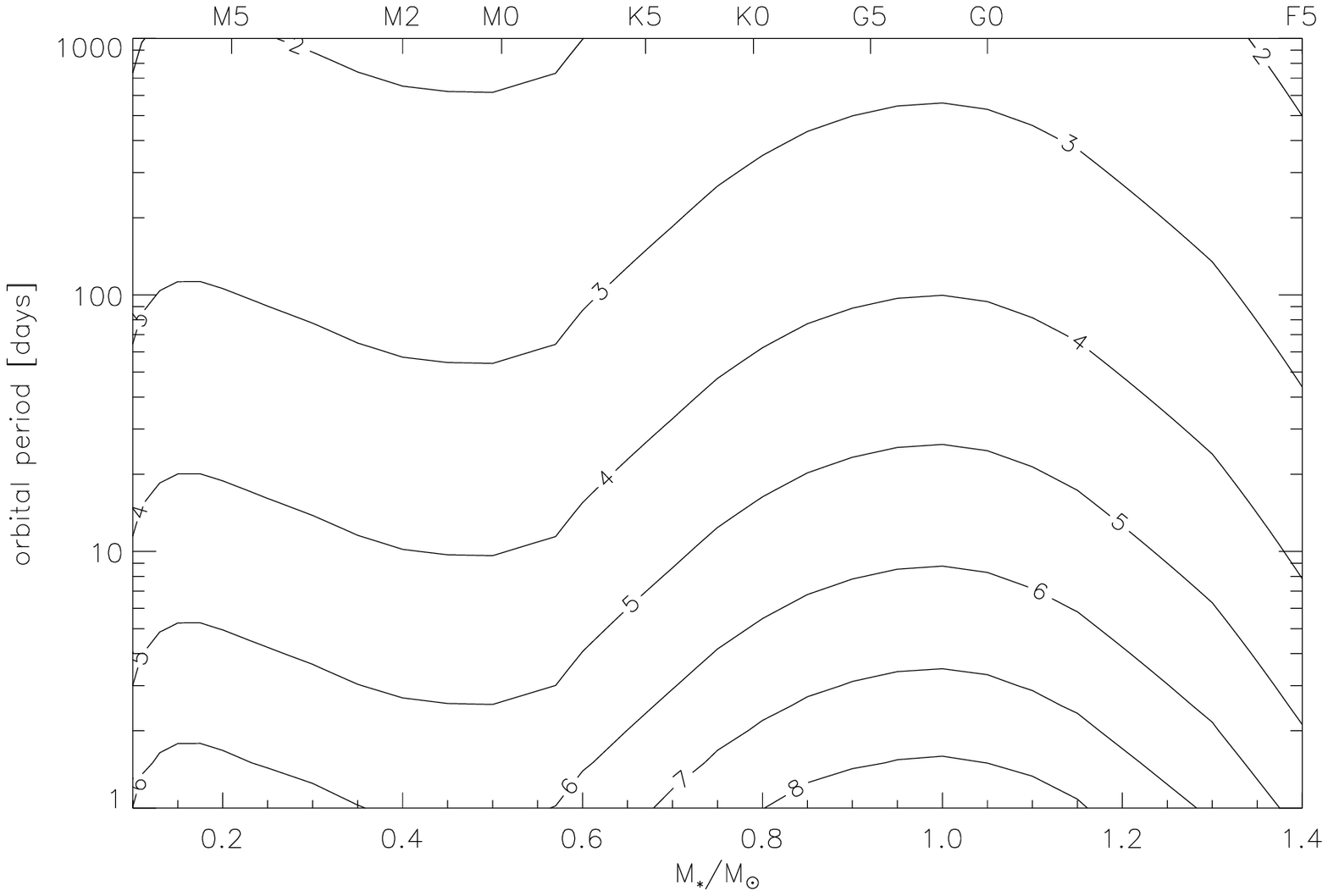}
  \caption{\label{f:prim-nep} S/N for a Neptune-size planet at 10\,pc, at the mid wavelength (3\,\um) and maximum average resolution of the considered \NIRSpec mode ($R$\,=\,100), and for a single primary transit. \textbf{For this figure and all following gas giant planet plots:} note the extended orbital period scale (where applicable).}
  \includegraphics[width=0.48\textwidth]{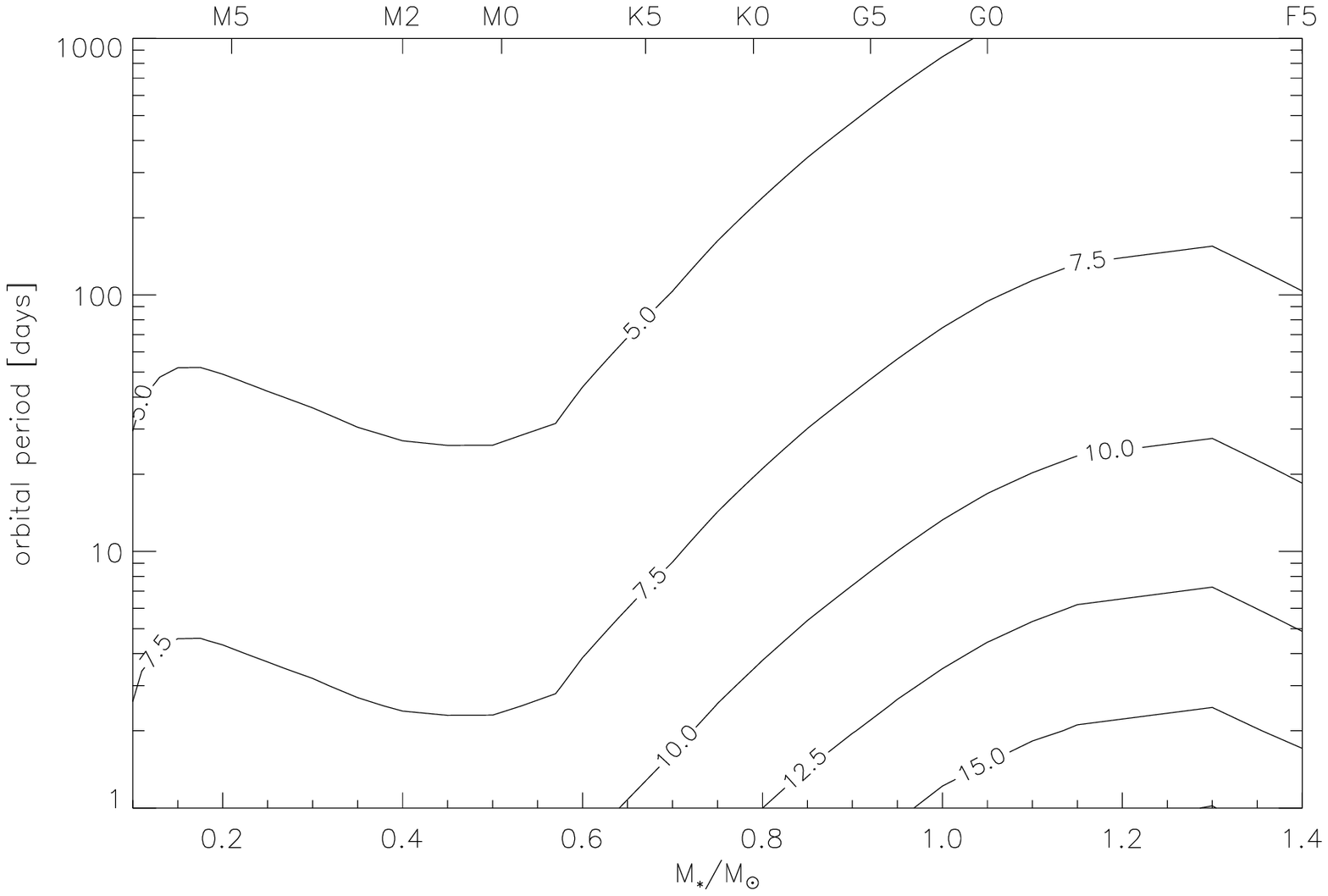} 
  \includegraphics[width=0.48\textwidth]{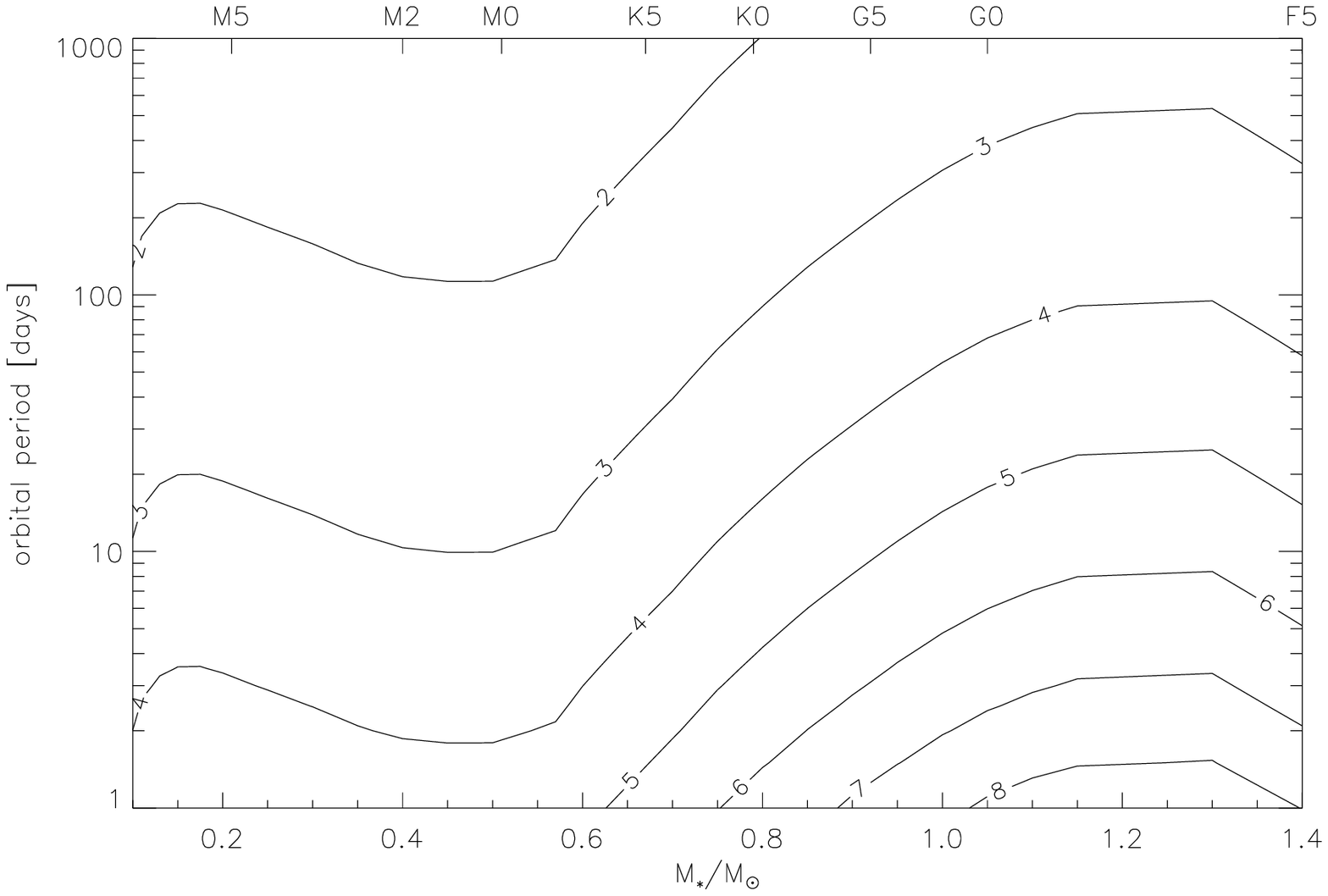}  
  \caption{\label{f:prim-jup} Same as Figure~\ref{f:prim-nep} for a Jupiter-size planet \textbf{at 50\,pc}.}  
\end{figure*}

Figure~\ref{f:prim-nep} shows that Neptune-size planets will require several transits to achieve spectroscopy at the maximum resolution of the mode ($R$ = 100). For jupiters, the performance on their S/N at 50\,pc (Figure~\ref{f:prim-jup}) is similar to that of the neptunes at 10\,pc above. However, given the greater distance, the saturation-induced curbing of the S/N with star mass is shifted  towards brighter stars: therefore Jupiter-size planets can be  characterized (at $R\,=\,100$) with only one transit up to a period of a few hundred days.

\subsubsection{MIRI}
\label{s:miri-first}

\begin{figure*}
  \centering
  \includegraphics[width=0.49\textwidth]{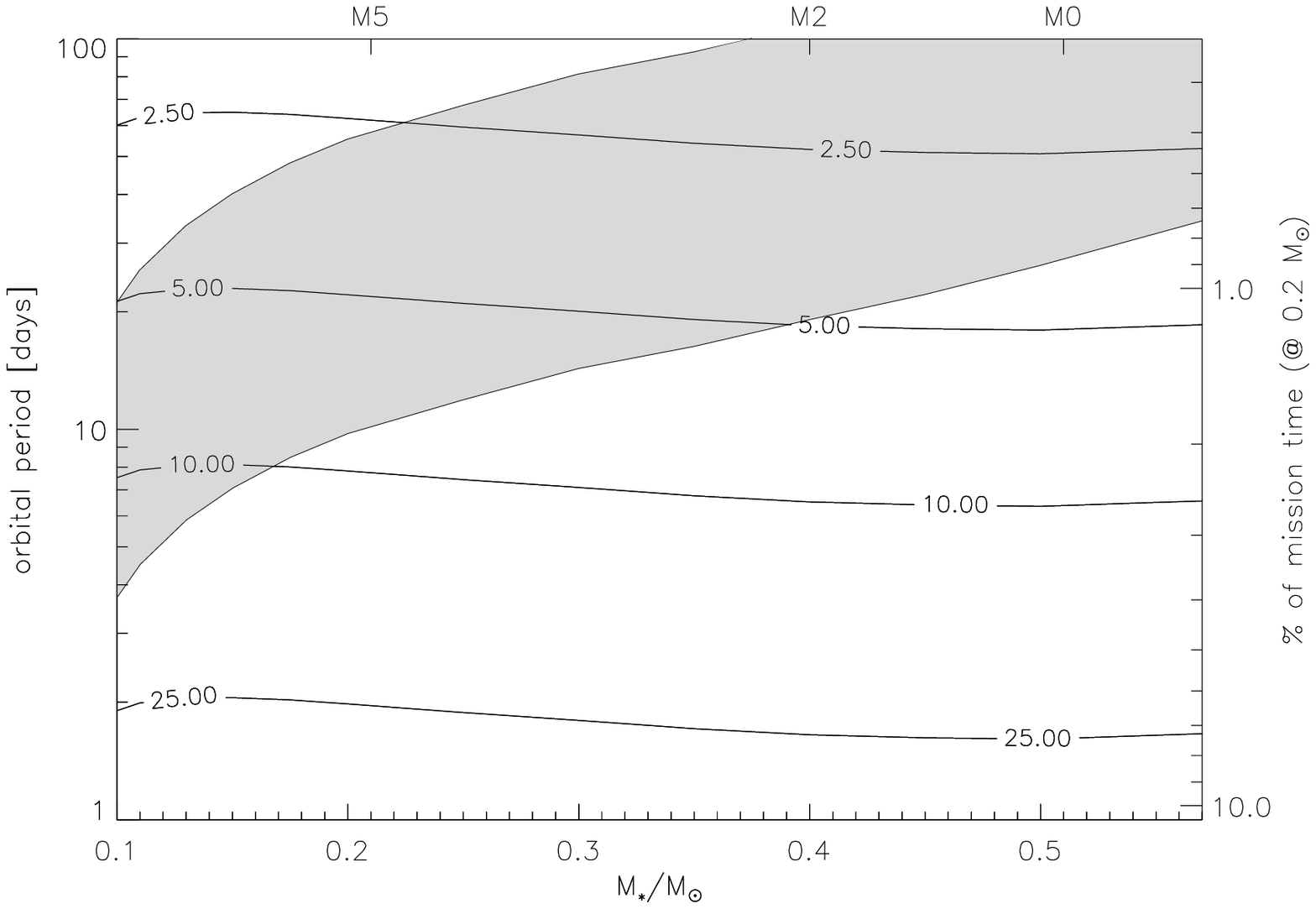}
  \includegraphics[width=0.49\textwidth]{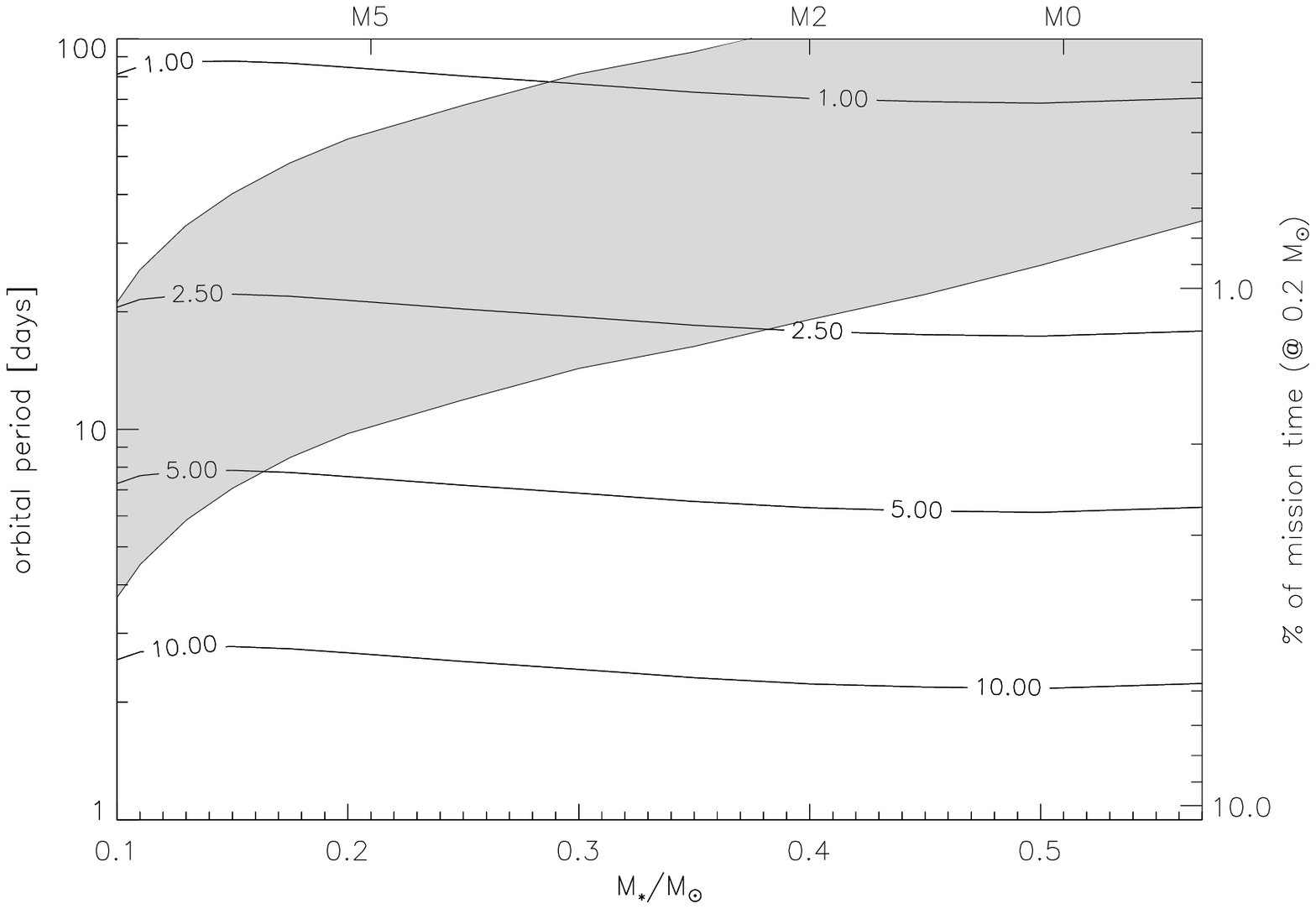}
  \caption{\label{f:O3-primary}S/N for the O$_3$ feature at 9.6\,\um{}, with the \MIRI instrument, in transmission. \textbf{For this plot and all subsequent O$_3$ ones:} a)\,$d\,=\,6.7\,\RM{pc}$, b)\,note the reduced star-mass range (close-up on the habitable zone).}
\end{figure*}

Figure~\ref{f:O3-primary} shows the S/N for the detection of the 9.6\,\um{} O$_3$ feature in our super-Earth prototype, for observation of all the primary transits available on average over the 5 year mission time, with stellar noise only (left) and all the above \MIRI and zodiacal noises (right). As seen above for the NIR spectroscopy, habitable planets can be characterized only around low mass stars. Additionally, because the O$_3$ feature is difficult to detect, we calculate this and all subsequent plots for this species at 6.7\,pc (value derived from occurring statistics for transiting habitable planets, see Section~\ref{s:stats}). It can be seen that the ozone feature will be detectable in transmission only for warm habitable planets around the lowest mass M dwarves. 

It should be noted that the chosen value for $\mu$ (18\,g\,mol$^{-1}$) implies an atmosphere dominated by  
water-vapor, corresponding to a super-Earth with a sufficient water reservoir near the inner edge of the habitable 
zone. O$_3$ detection   
as well as its interpretation in terms of biosignature are problematic  
within a H$_2$O-rich atmosphere:
\begin{itemize}
	\item photochemical productions of H, OH, HO$_x$ associated with H$_2$O  
photolysis drastically limit the build-up of an ozone layer \citep{selsis_2002},
	\item collision broadening far-wing absorption by H$_2$O can screen the O$_3$ signature (see Fig. 4,  
\citealt{selsis_screening}),
	\item the presence of H$_2$O above the tropopause yields enhanced loss of  
hydrogen to space and the abiotic build-up of oxygen leftovers, making  
the indirect biological origin of O$_3$ doubtful \citep{sels_gl581}.
\end{itemize}

\begin{figure*}
  \centering  
  \includegraphics[width=0.49\textwidth]{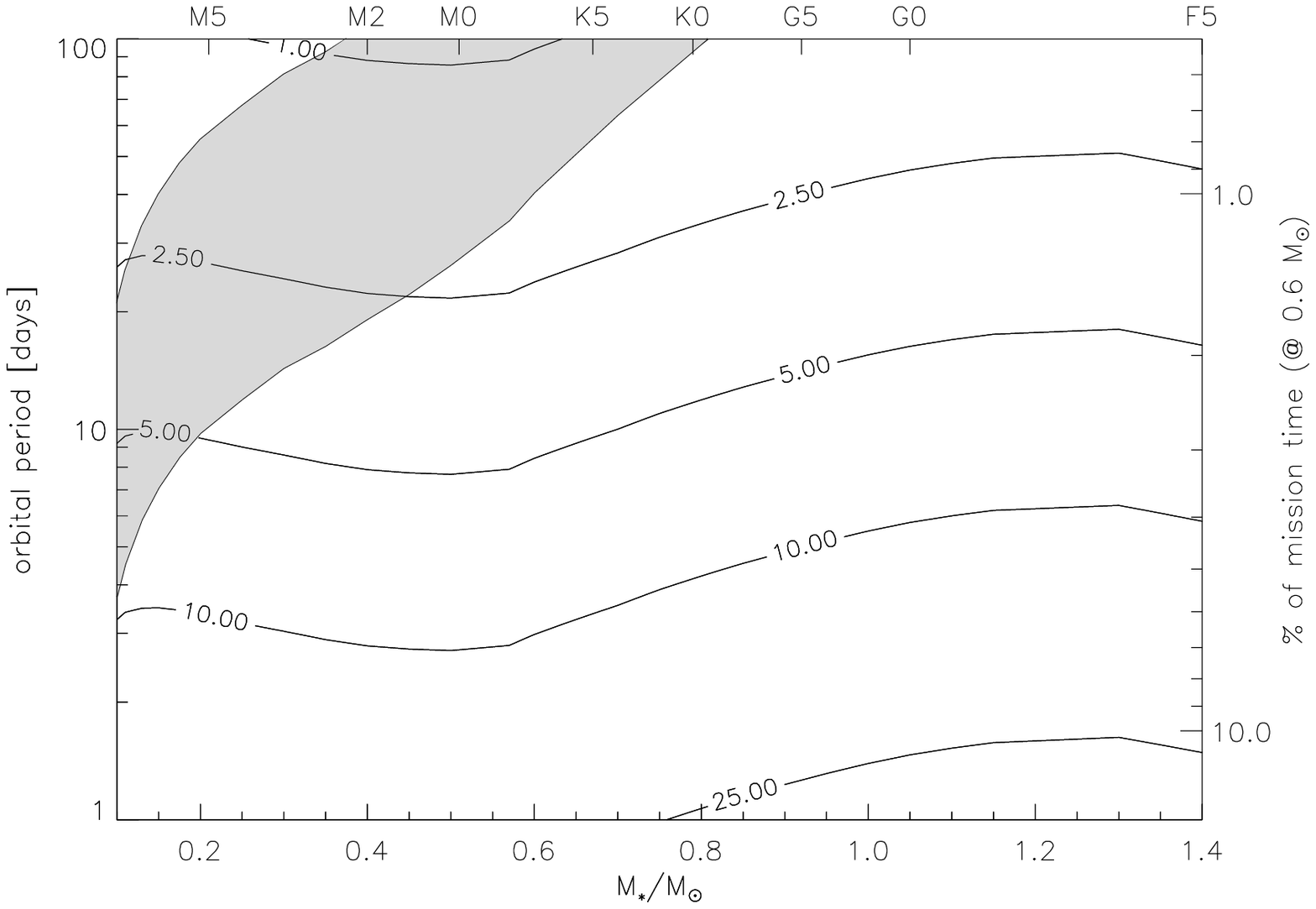}
  \includegraphics[width=0.49\textwidth]{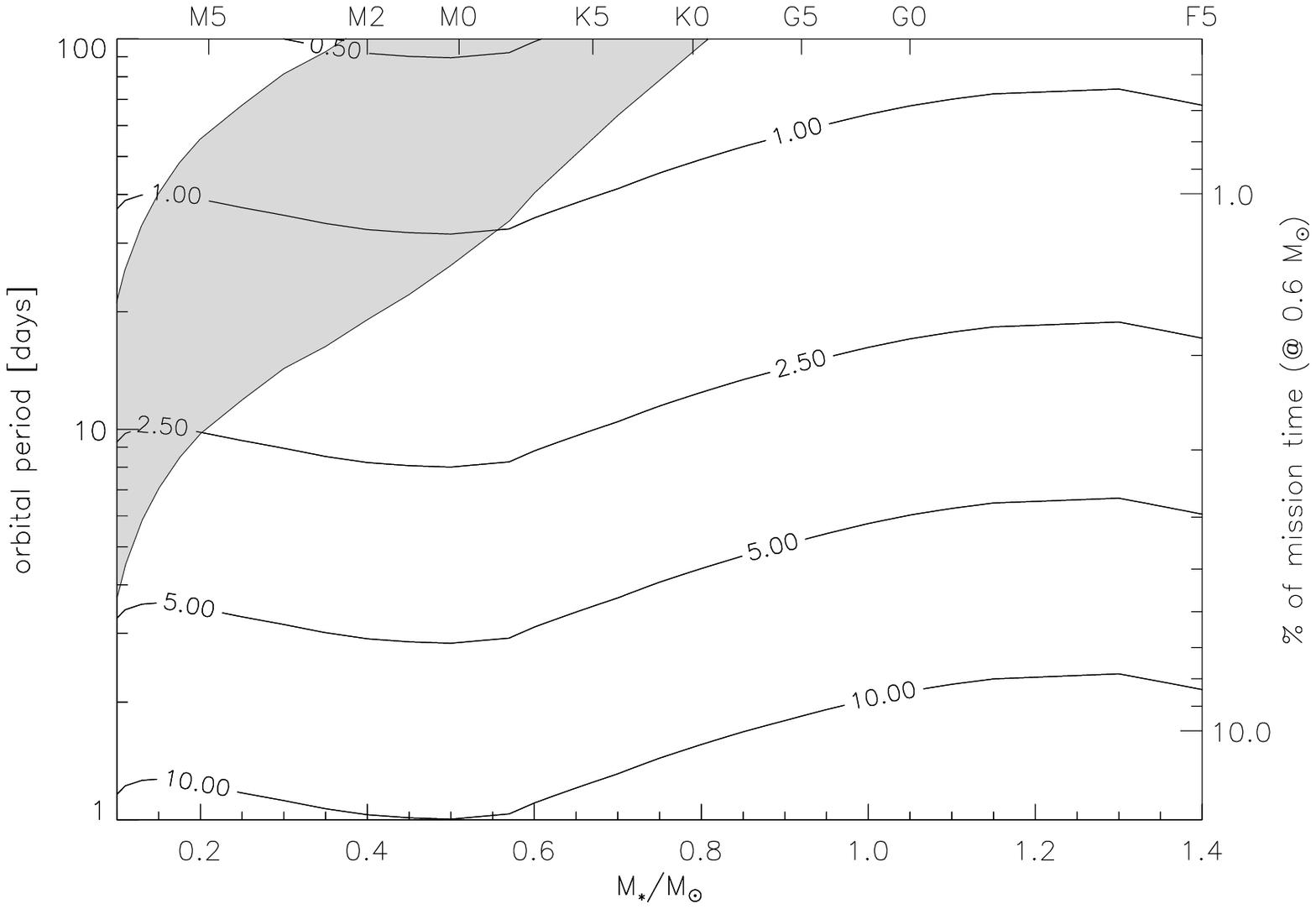}
  \caption{\label{f:CO2-primary}Same as Figure~\ref{f:O3-primary}, but for the 15\,\um{} CO$_2$ feature, and at 10\,pc.}
\end{figure*}

\begin{figure*}
  \centering  
  \includegraphics[width=0.49\textwidth]{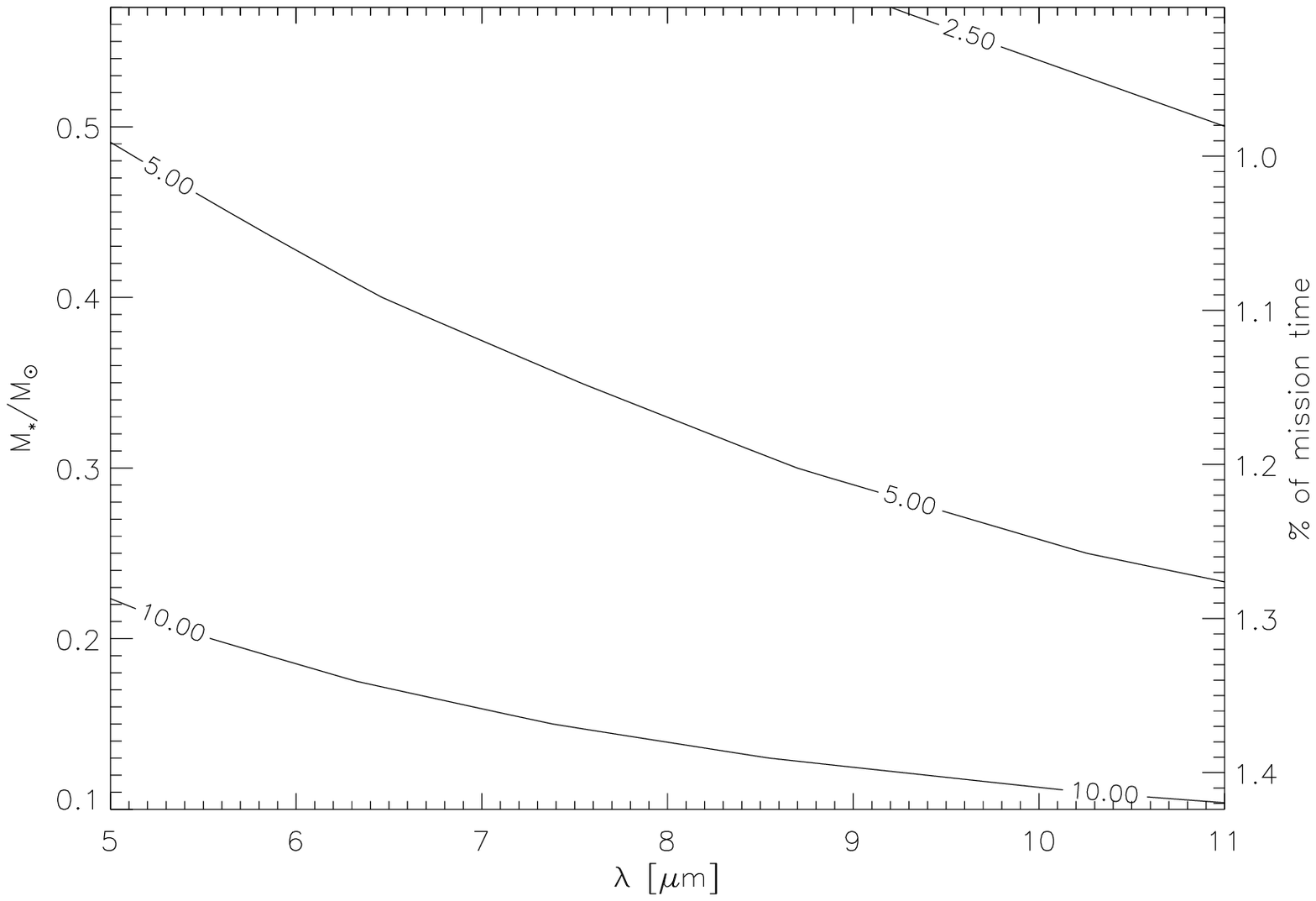}
  \includegraphics[width=0.49\textwidth]{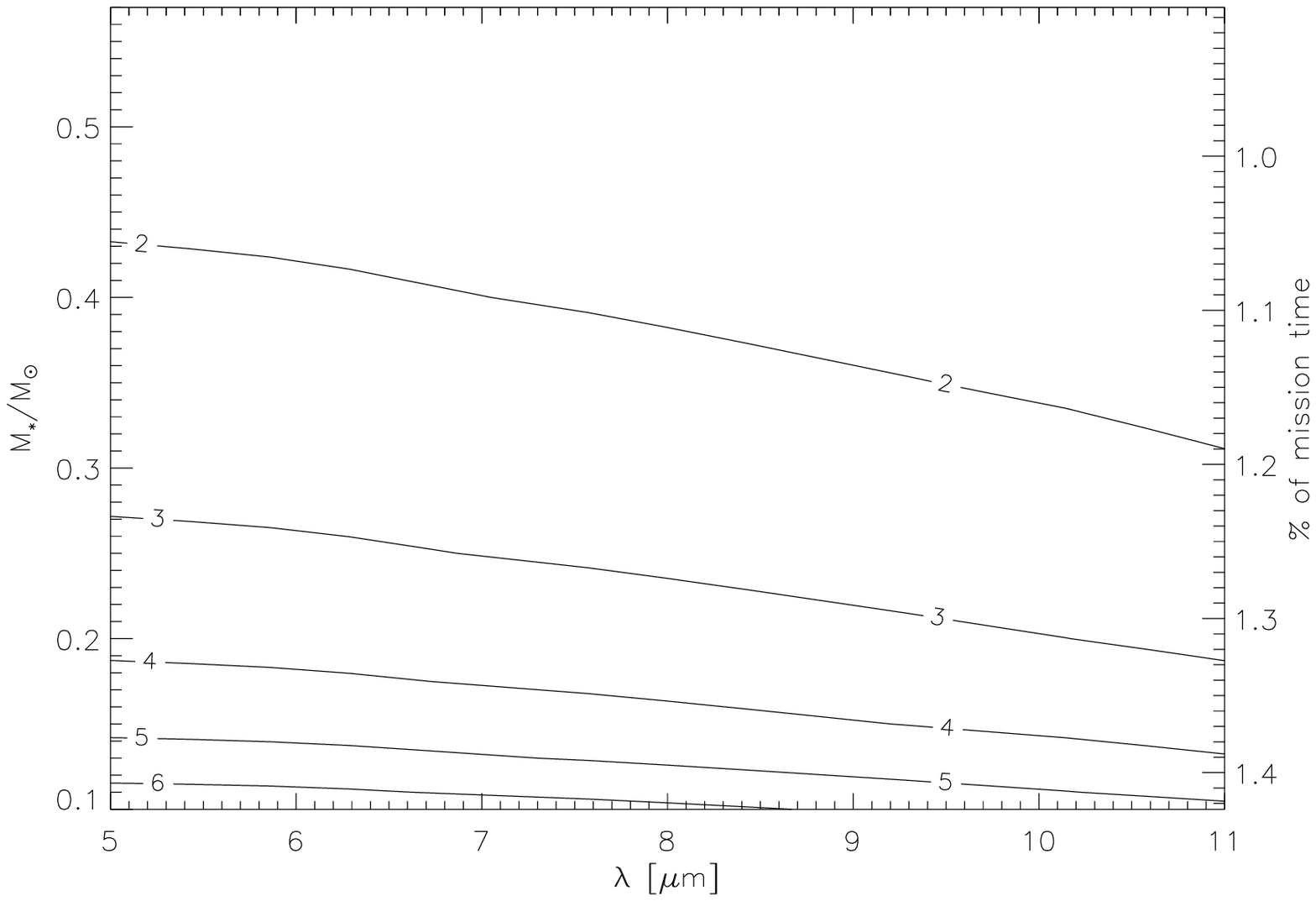}
  \caption{\label{f:hz1-miri-primary}Primary transit full spectroscopy S/N for a super-Earth at 10\,pc with \MIRI, in the (wavelength - star mass) parameter space. The planet's distance to the star is such that it receives the same amount of energy as Earth (1 AU from the Sun). The width of the fiducial spectral feature used in the computation corresponds at each wavelength to a fixed resolution ($R$=20).}
\end{figure*}

In order to compare with forthcoming work from Cavarroc et al., we also consider the CO$_2$ feature at 15\,\um. We use the filter \textsl{IM\_6}, 3\,\um{} wide, centered at 15\,\um. No saturation occurs for this observation for the range of considered stellar masses, assuming that we can read only the PSF window (2 times the size of the PSF corresponding to the upper wavelength bound of the redmost filter (\textsl{IM\_6}). For the out-of-feature reference, we use the \textsl{IM\_4} filter, running from 10.95 to 11.65\,\um, that is 3 times narrower than \textsl{IM\_6}. We therefore substitute the factor 4 in Eq.~\ref{eq:s2n} by $\left(^3/_9+1\right)+2\approx3.33$, but only for the bandwidth-dependent terms (i.e stellar, zodi and thermal). The effective bandwidth considered is therefore that of the out-of-feature filter (\textsl{IM\_4}, 0.7\,\um), and the effective wavelength is $^{11+15}/_2 = 13$\,\um.  Figure~\ref{f:CO2-primary} shows the S/N for this feature. Given the uncertainties in our model, we can see that CO$_2$ at 15\,\um{} will be difficult to detect in transmission even for warm habitable planets around the lowest mass stars.

At constant stellar mass, the reduction in S/N is a composite of a)\,the uniform effect of the readout noise, b)\, the reduced quantum efficiency of the MIRI detector for the shorter wavelengths, c)\, the thermal and the local zodiacal contribution towards longer wavelengths (Figure~\ref{f:hz1-miri-primary}). It can be seen that performance increases towards shorter wavelengths, where additional signatures such as methane and water bands are situated.

\begin{figure*}
  \centering
  \includegraphics[width=0.49\textwidth]{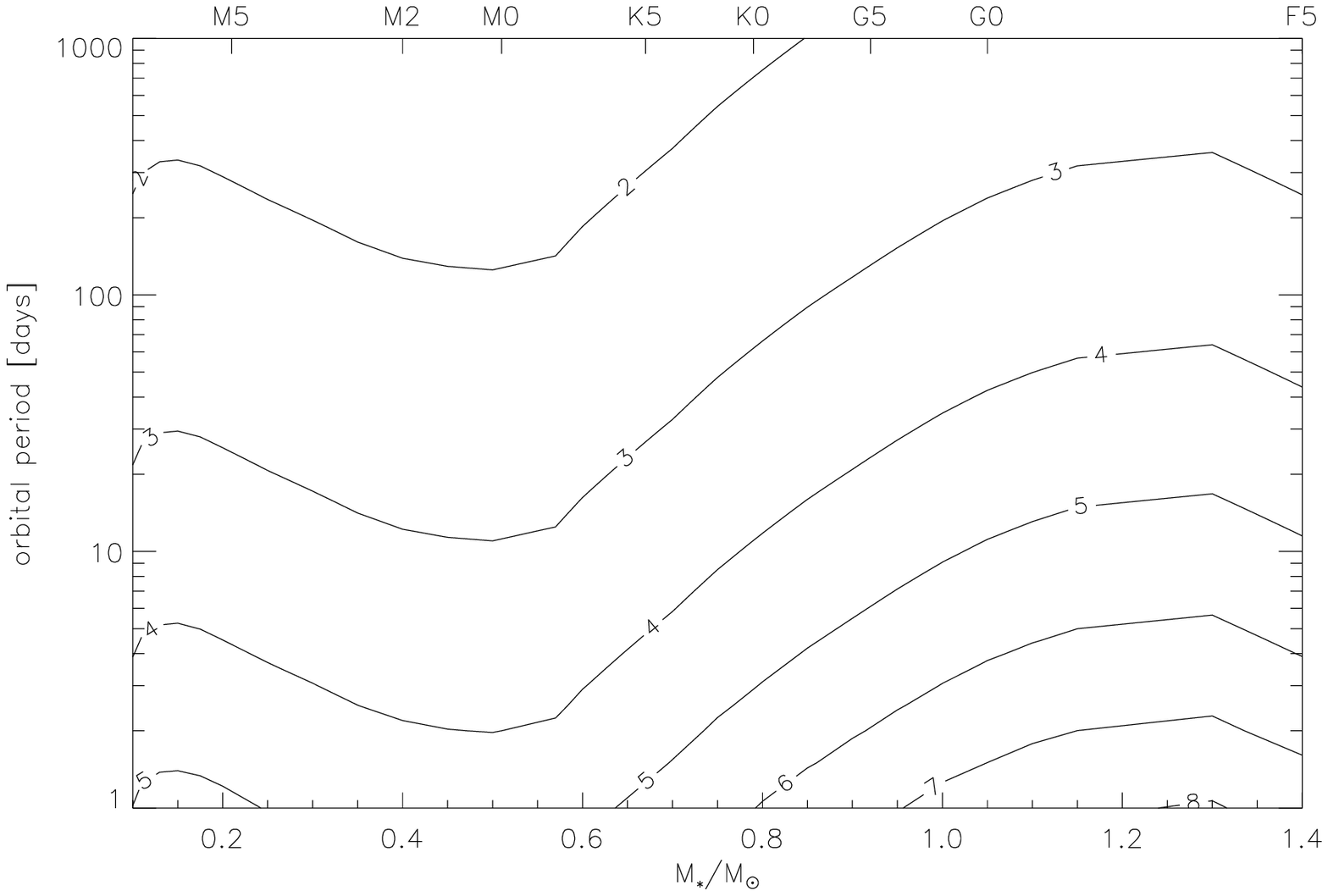} 
  \includegraphics[width=0.49\textwidth]{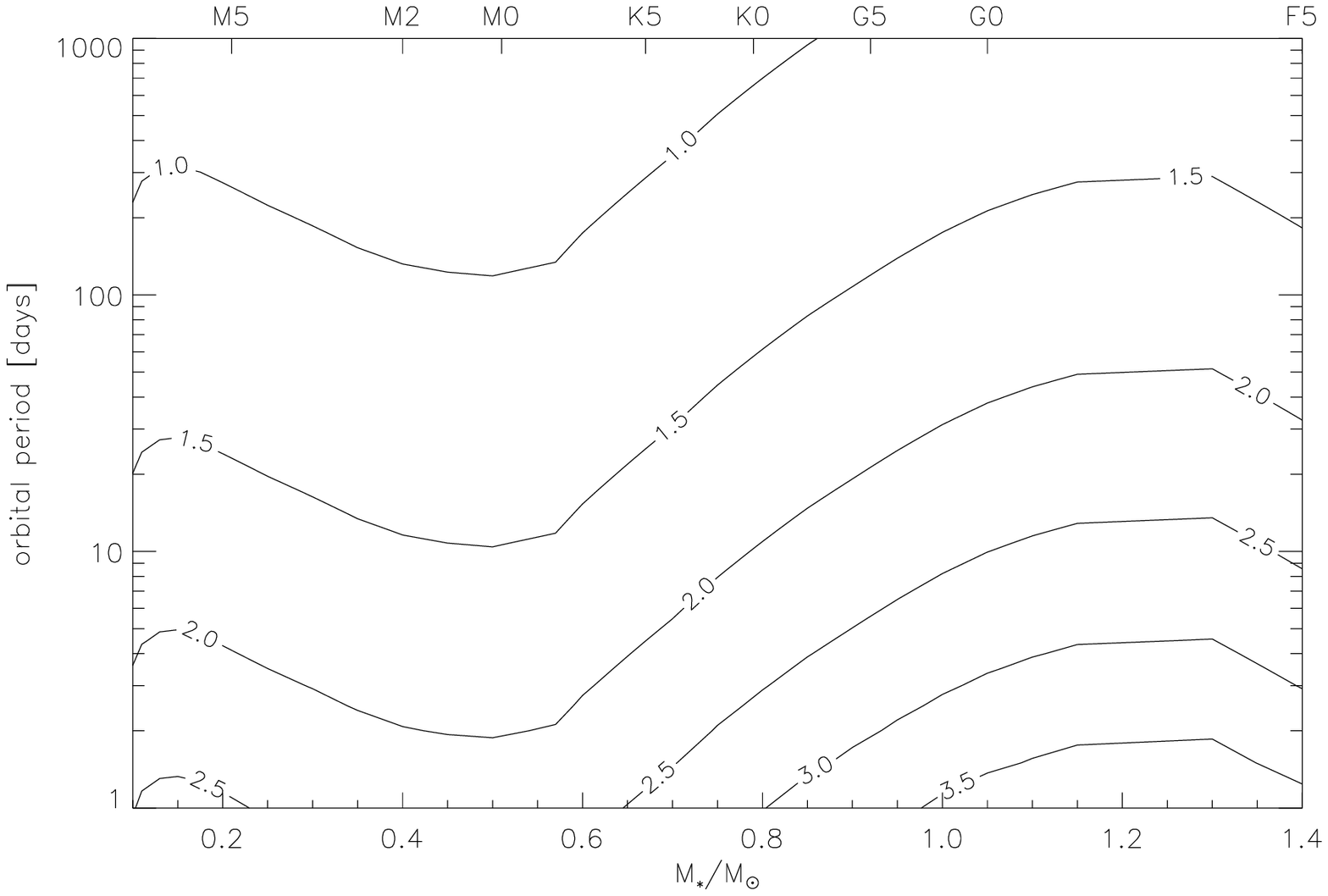}
  \caption{\label{f:nep-miri-10um-primary}S/N for a Neptune-size planet, for a single primary transit, at 10\,\um{} (\MIRI), and at the full resolution ($R\,=\,100$) of the instrument.}
\end{figure*}

\begin{figure*}
  \centering
  \includegraphics[width=0.49\textwidth]{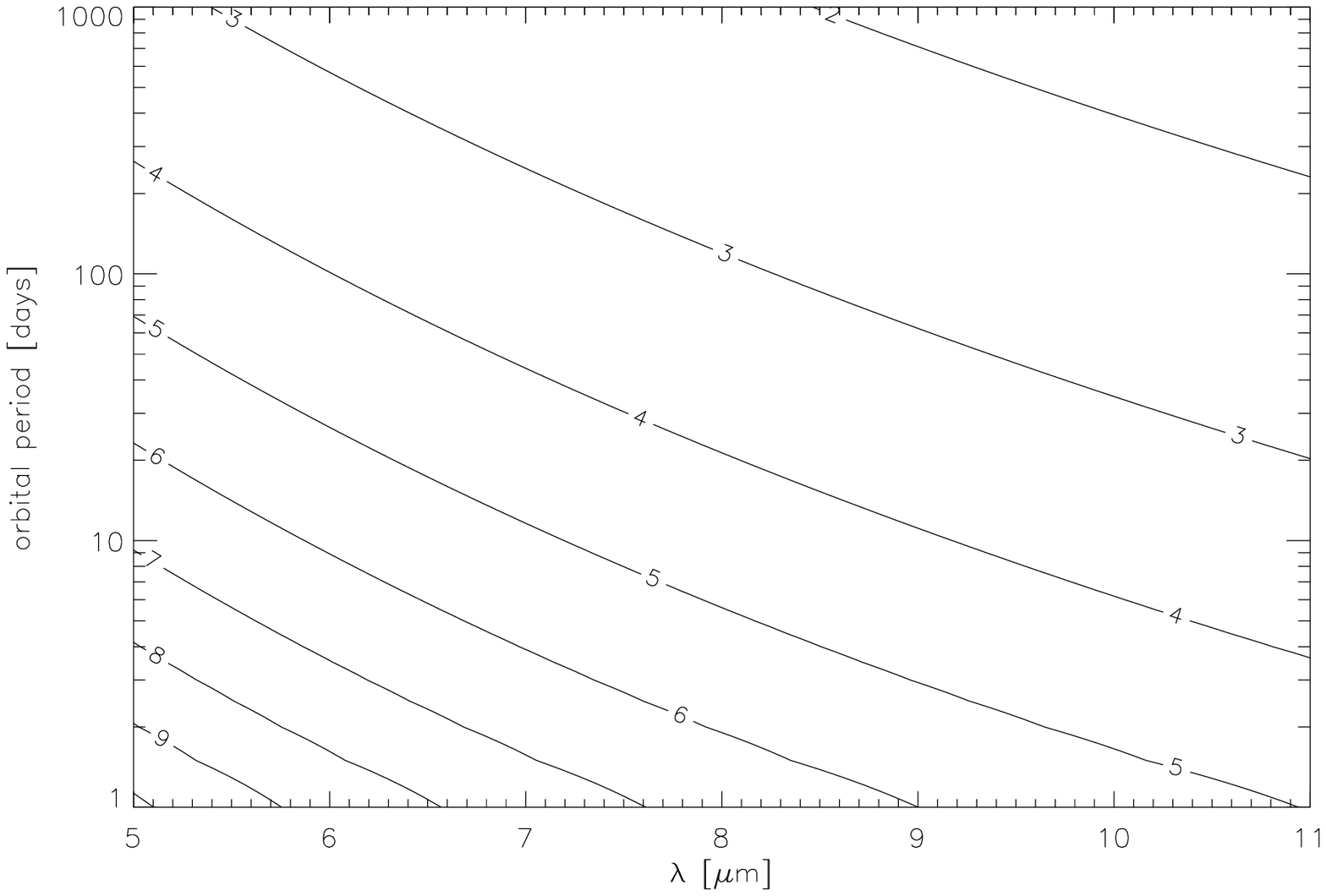}
  \includegraphics[width=0.49\textwidth]{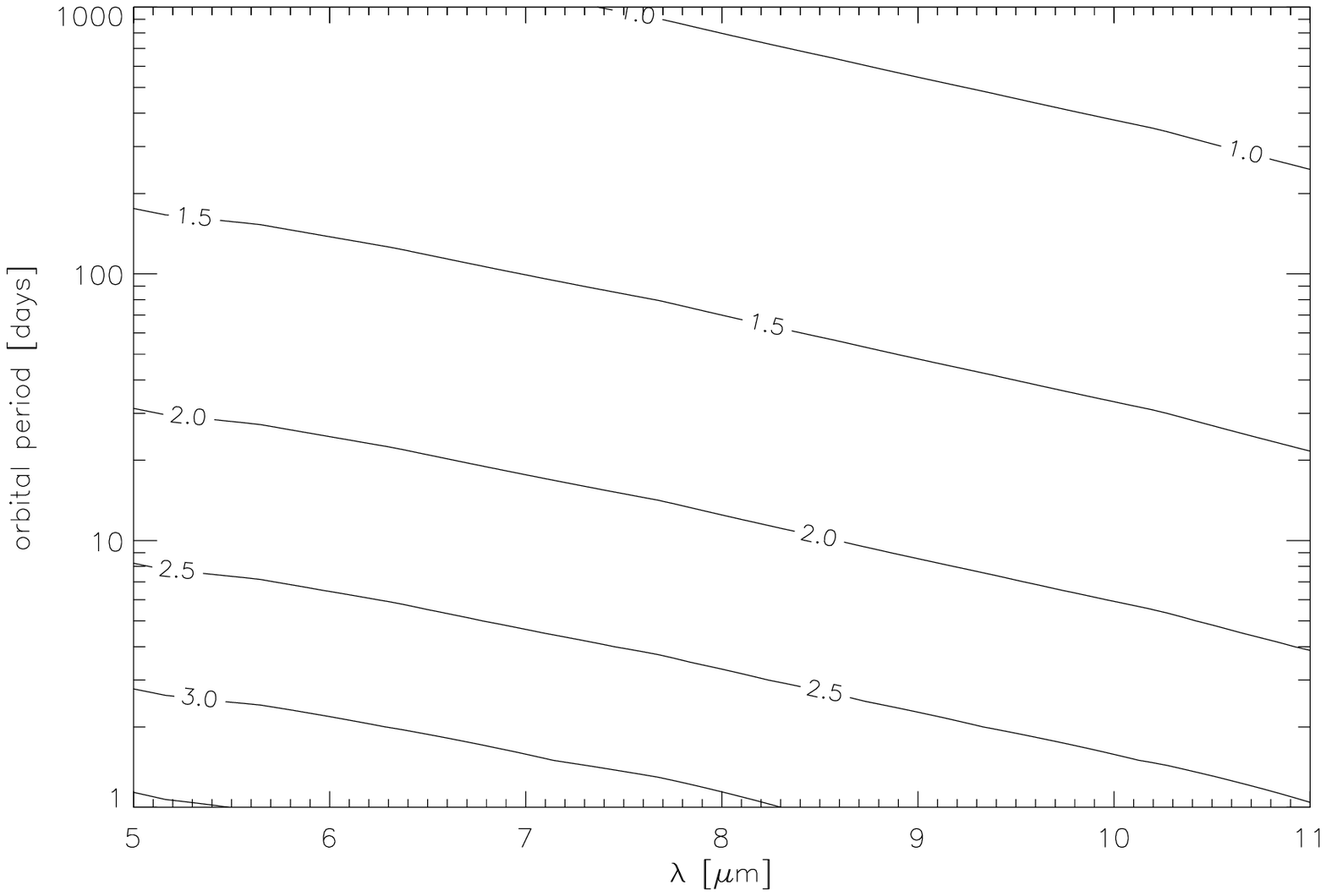}
  \caption{\label{f:primary-jup-miri-Msun}Same as Figure~\ref{f:nep-miri-10um-primary}, but for a Jupiter-size planet \textbf{at 50 pc} around a solar-mass star, in the wavelength-planet period space.}
\end{figure*}

Thanks to their lower density\footnote{Thus both high radius and high $H$.}, because of their high hydrogen content, Neptune-size planets can be spectroscopically characterized from a single or a couple of transit observations. Figure~\ref{f:nep-miri-10um-primary} shows the S/N \emph{for a single transit} of our prototype Neptune planet, towards the most unfavorable end of the \MIRI's \LRS wavelength range. The S/N for our Jupiter prototype in the (wavelength-planet period) space, around a solar type star, and at the full resolution of the instrument, is presented in Figure~\ref{f:primary-jup-miri-Msun}.\\
\\
\\
It can be noted that for the giant planets, for \NIRSpec as well as for \MIRI, the dynamic of the S/N over the parameter space is not very strong, implying that these planets can be characterized in a wide variety of cases with a fairly constant number of cumulated transits.

\subsection{Secondary transit - emission}

\New{It is interesting to note that the atmospheric species detectable in an emission spectrum are, by definition, greenhouse gases, which affect the planet's climate.}

\subsubsection{NIRSpec}

\begin{figure*}
  \centering
  \includegraphics[width=0.49\textwidth]{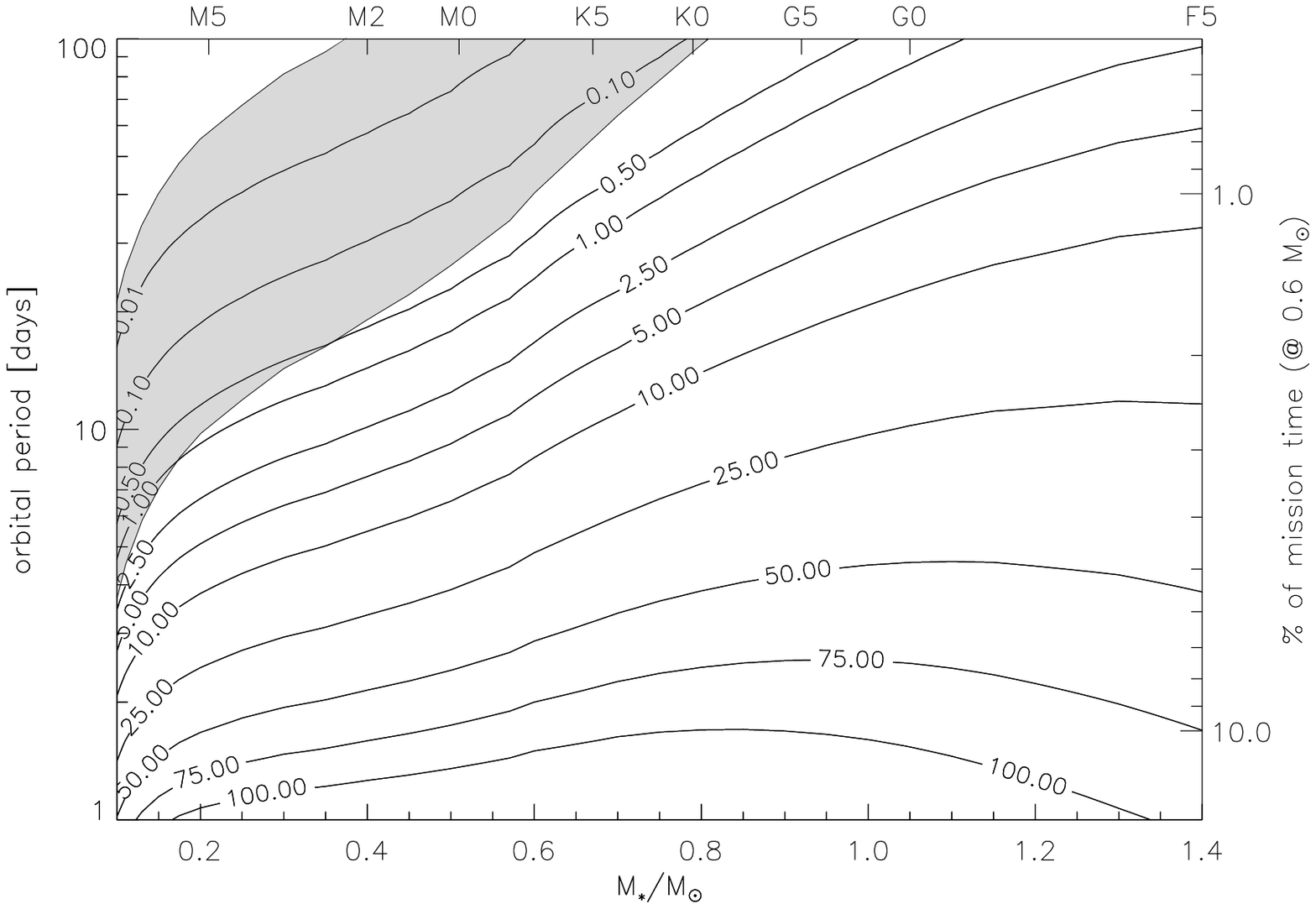}
  \includegraphics[width=0.49\textwidth]{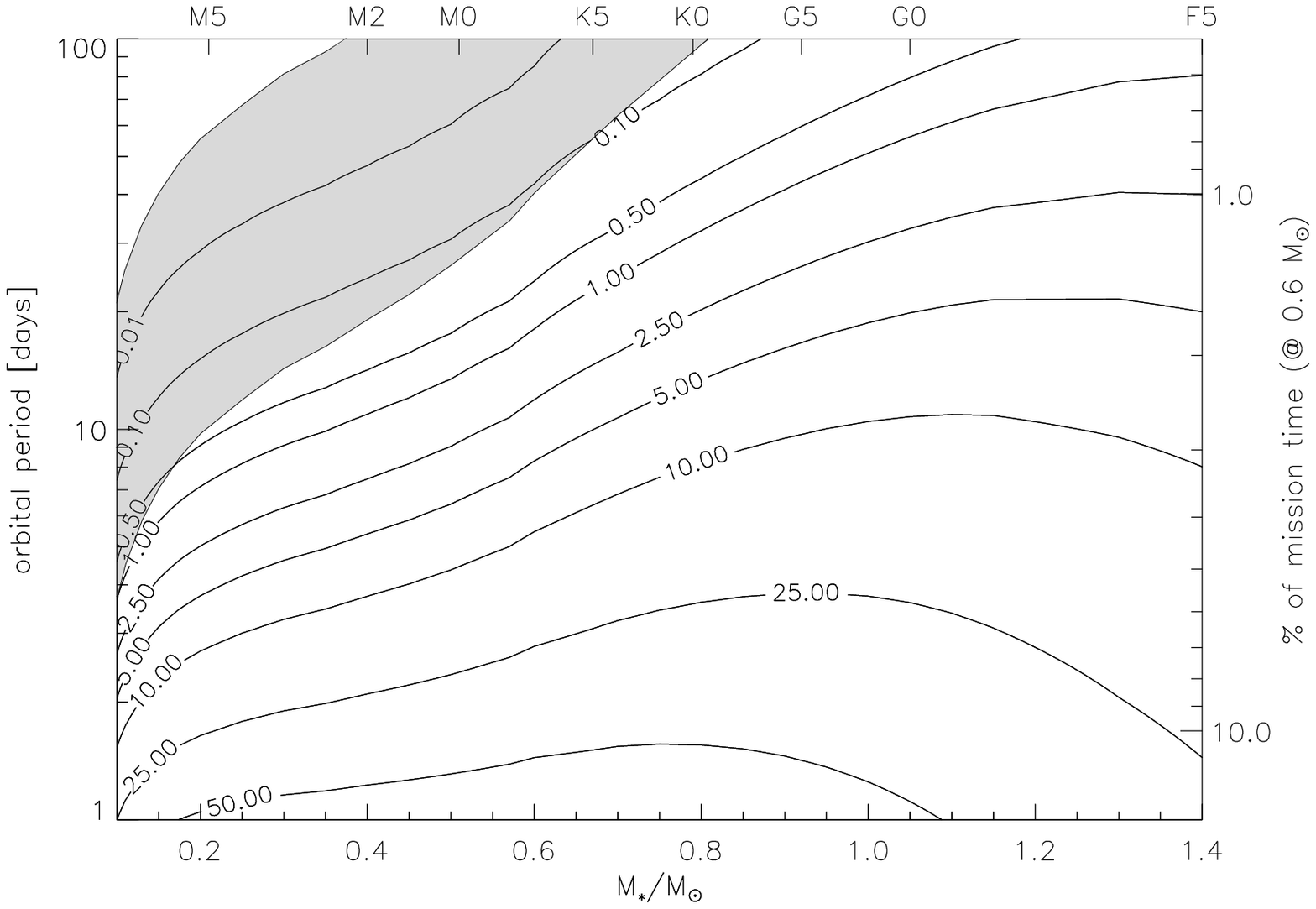}

  \caption{\label{f:emiss-CO2-searth}S/N in emission of the 4.3\,\um{} CO$_2$ signature in a super-Earth at 10\,pc, for observations of all secondary transits available on average over the 5 year mission time, with stellar noise only (\textbf{left}) and modeled NIRSpec and zodiacal noises (\textbf{right}).}
\end{figure*}

Even when integrating over the whole mission time, the 4.3\,\um{} CO$_2$ feature is not detectable in emission on habitable super-earths, even those on the inner edge of the habitable zone (Figure~\ref{f:emiss-CO2-searth}). However, the CO$_2$ signature can be detected in emission on hot super-Earths around low mass stars. Let us suppose for instance a high gravity planet and a dense atmosphere, that would lower the scale height and prevent detection in transmission. D	etection of CO$_2$ through emission would be an indicator of the presence of an atmosphere, which is a question of debate for these low mass objects at short distances from active stars.

\begin{figure*}
  \centering
  \includegraphics[width=0.49\textwidth]{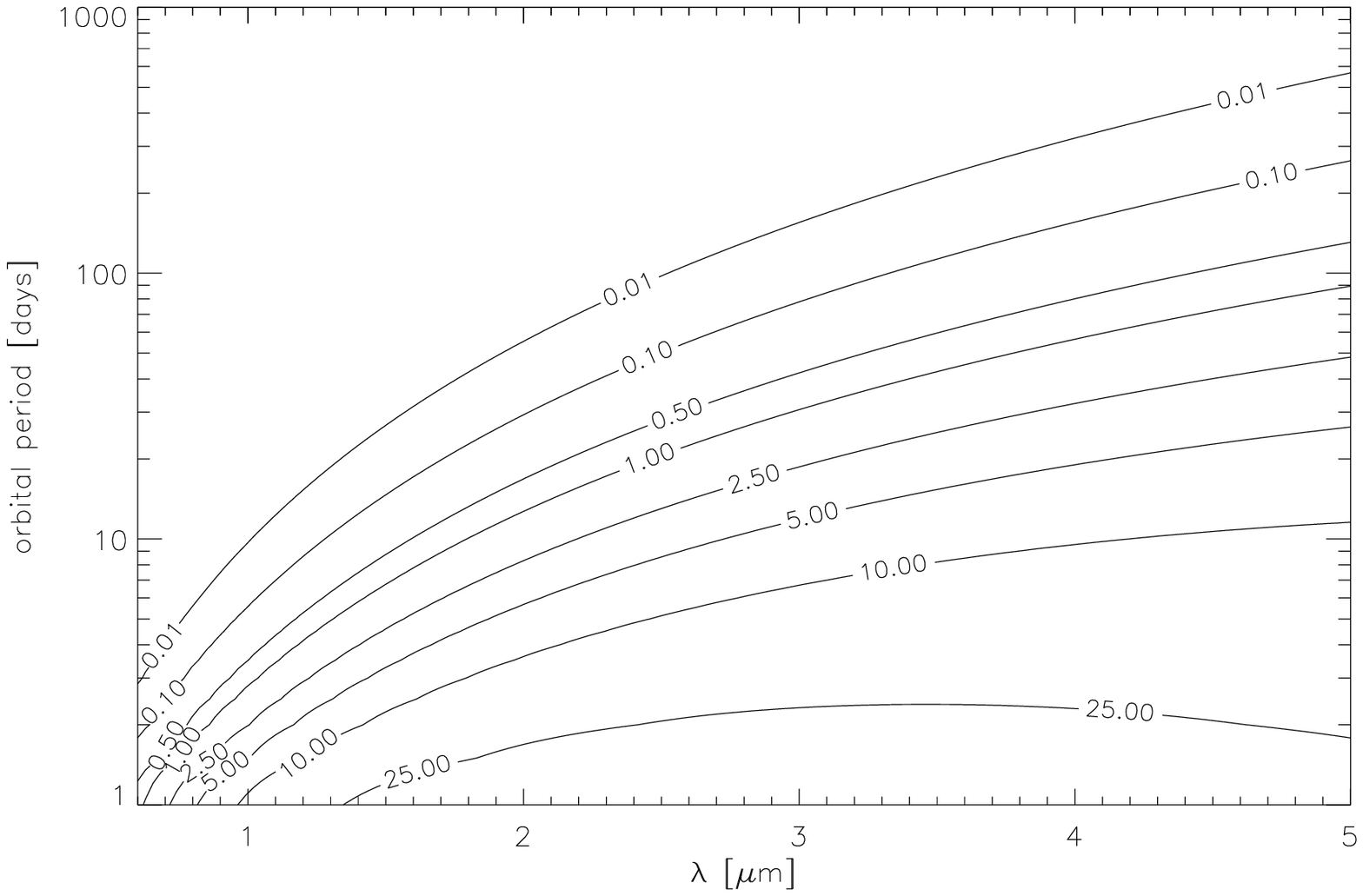}
  \includegraphics[width=0.49\textwidth]{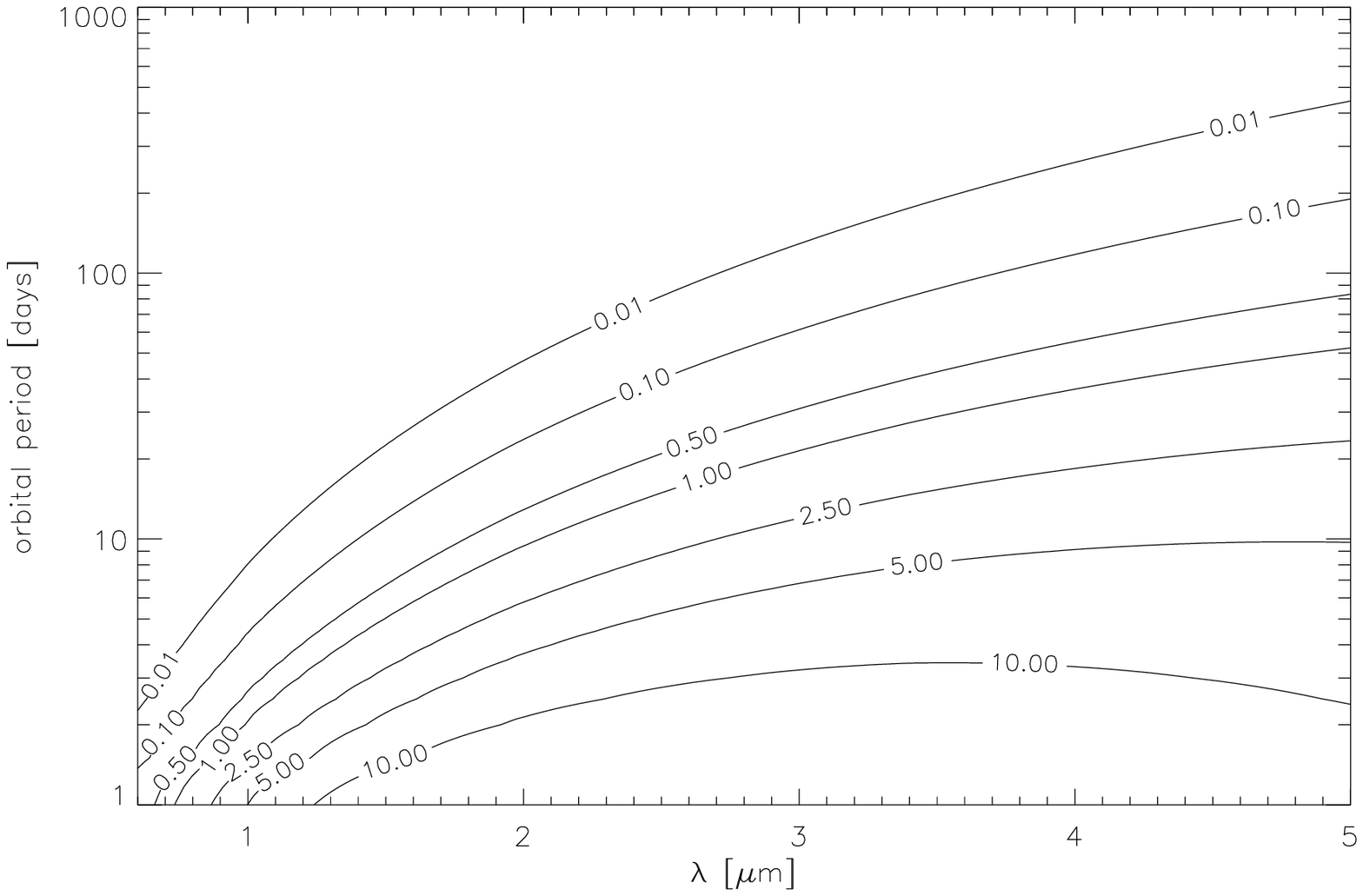}

  \caption{\label{f:emiss-Jup-NIR-Msun}Same as Figure~\ref{f:emiss-CO2-searth} but for a single secondary transit of a Jupiter-size planet, \textbf{at 50 pc}, around a solar type star, in the wavelength-planet period space, and at the full resolution of the NIRSpec $R\,=\,100$ mode. The depth of the fiducial spectral signature is $\alpha\,T_\RM{eq}$ with $\alpha = 0.2$.}
\end{figure*}

Unlike the case of transmission spectroscopy, Neptune-mass planets cannot be characterized in emission with \NIRSpec with only one or a couple of transits (plot not shown). It is required to go up to the Jupiter-mass scale to recover this ability, and only for the hot planets. The performance on the S/N drops drastically shortward 2\,\um{} (Figure~\ref{f:emiss-Jup-NIR-Msun}). For these cases, the strength in emission of our $R=100$-equivalent wide fiducial spectral feature is $\Delta \Tb\,=\, \alpha\,T_\RM{eq}$ with an optimistic $\alpha\,=\,0.2$.

\subsubsection{MIRI}

\begin{figure*}
  \centering
  \includegraphics[width=0.49\textwidth]{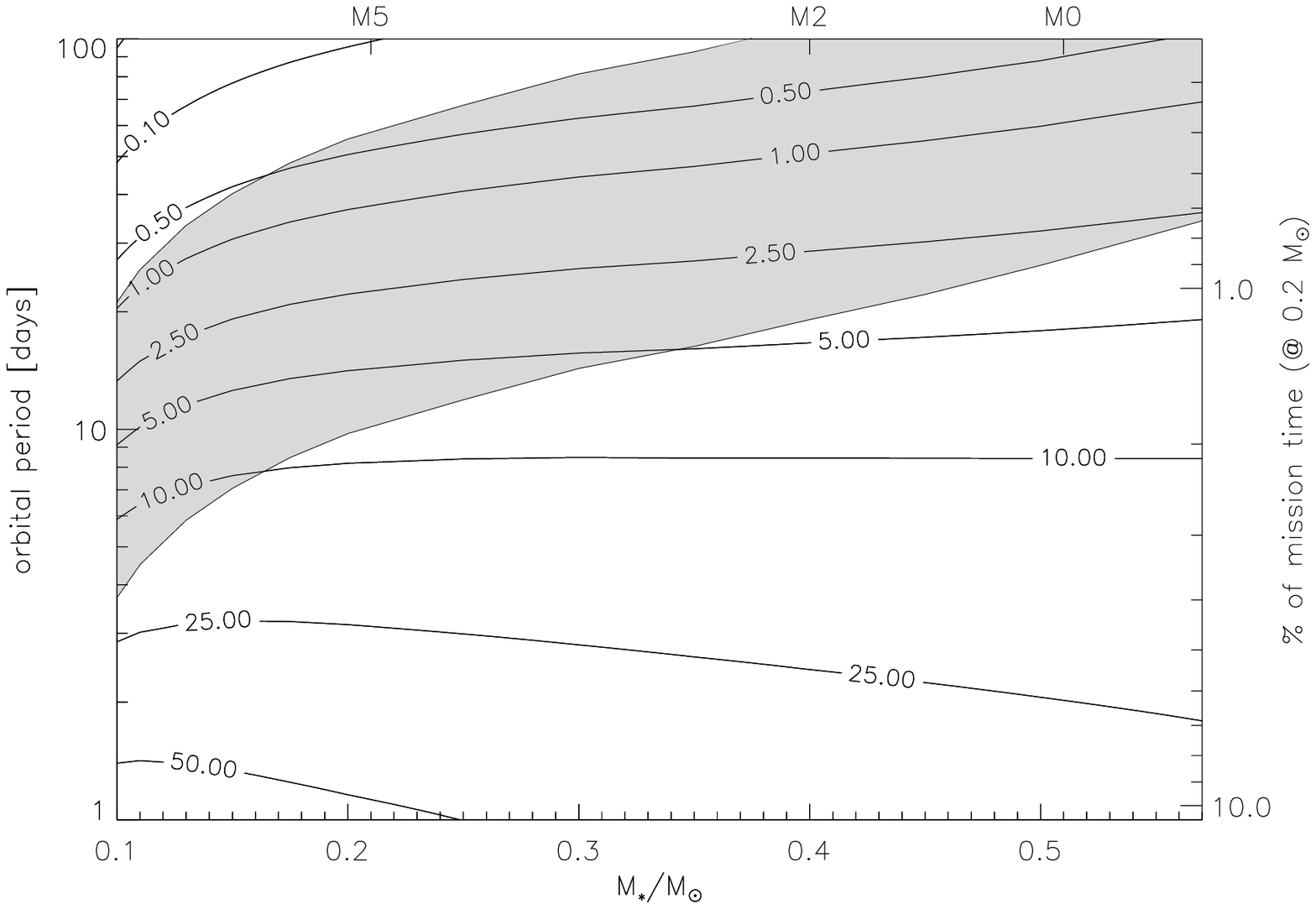}
  \includegraphics[width=0.49\textwidth]{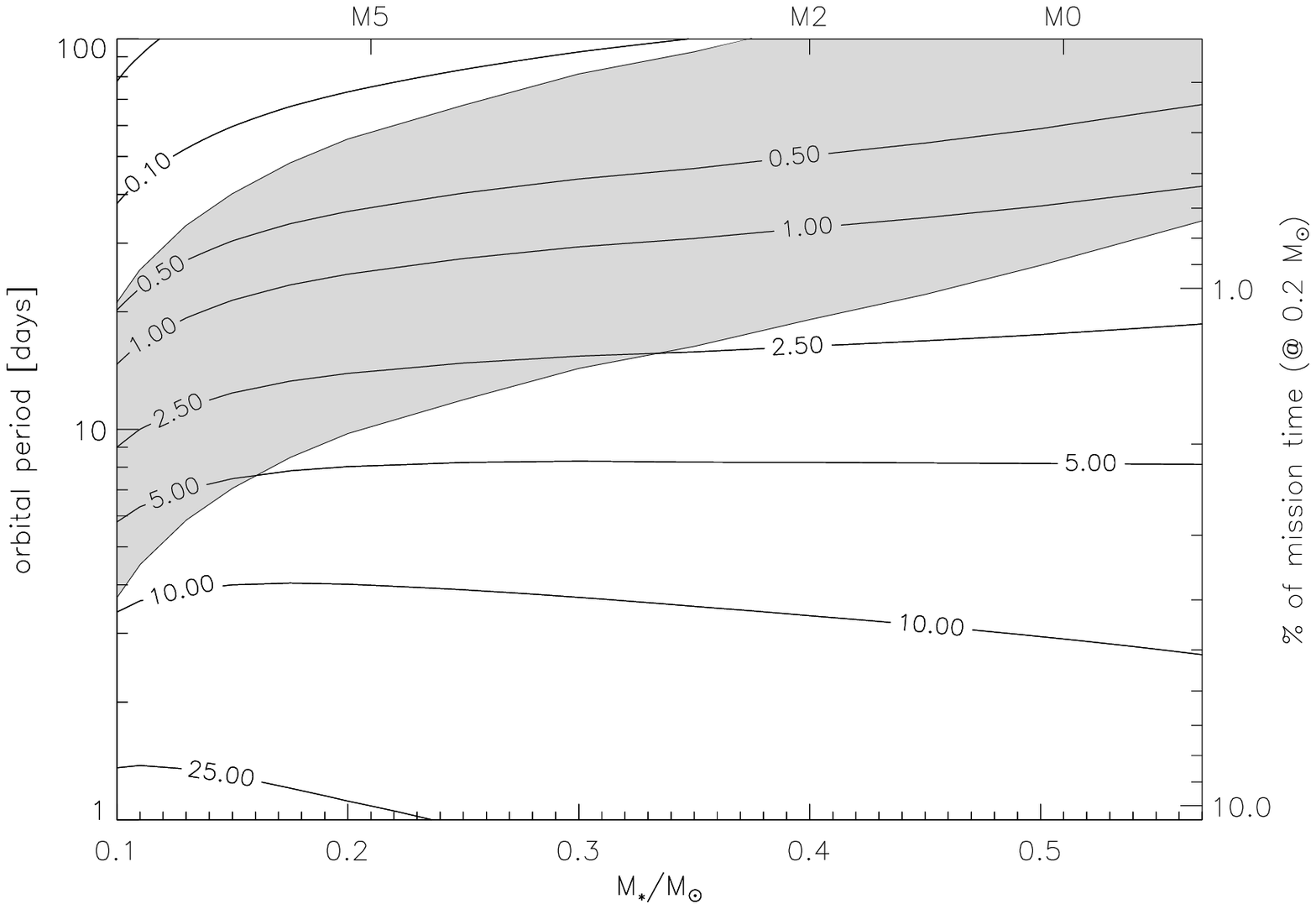}

  \caption{\label{f:emiss-O3}S/N in emission for the 9.6\,\um{} O$_3$ signature, with the \MIRI instrument for a star situated at \textbf{6.7\,pc}.}
\end{figure*}

Figure~\ref{f:emiss-O3} shows the secondary transit-emission spectroscopy, for our habitable planet prototype, for the O$_3$ feature. With our current set of parameters, the signature appears less detectable than in primary transit (Figure~\ref{f:O3-primary}).

\begin{figure*}
  \centering
  \includegraphics[width=0.49\textwidth]{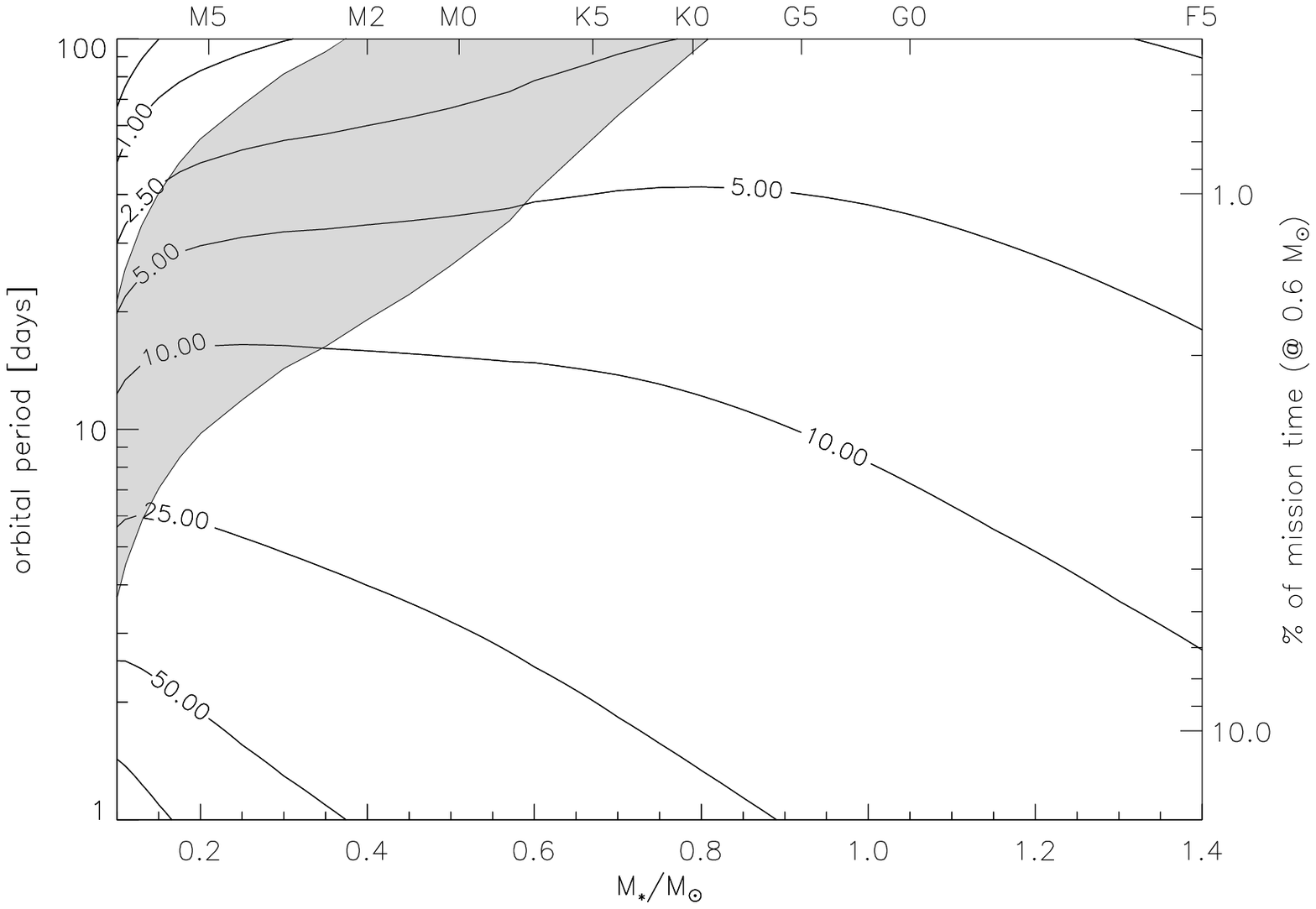}
  \includegraphics[width=0.49\textwidth]{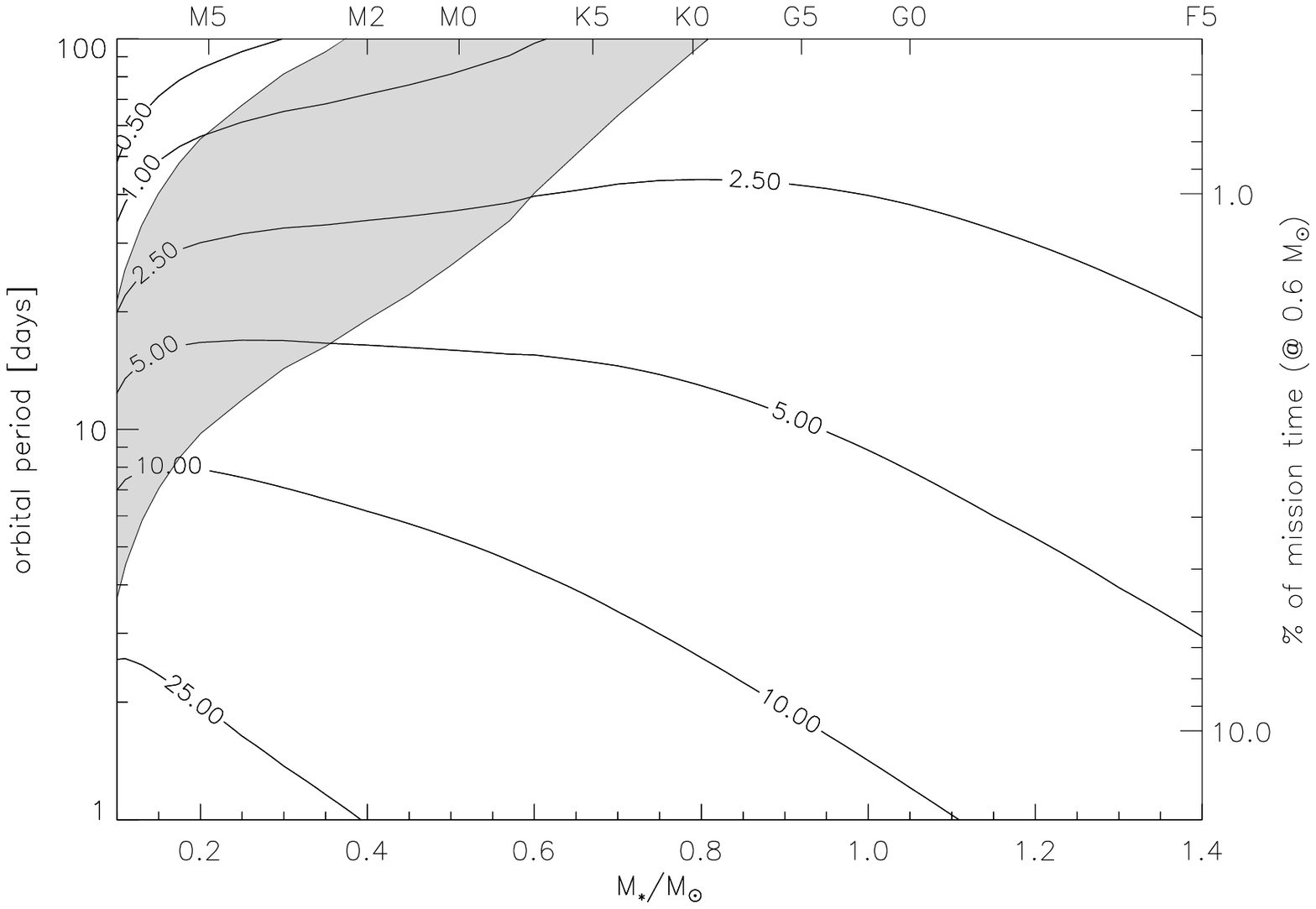}
  
  \caption{\label{f:15um-emission-CO2}Same as Figure~\ref{f:emiss-O3}, but for the CO$_2$ spectral feature at 15\,\um{}, and at 10\,pc.}
\end{figure*}

We examine then the 15\,\um{} CO$_2$ signature in emission, with the same modeling as in Section~\ref{s:miri-first}. Figure~\ref{f:15um-emission-CO2} shows that (given the considered parameters) the detection of CO$_2$ in emission at 15\,\um{} is a little less efficient than in transmission at 4.3\,\um{} (Figure~\ref{f:4.3-CO2-trans}).

\begin{figure*}
  \centering 
  $\begin{array}{cc}   
    \includegraphics[width=0.49\textwidth]{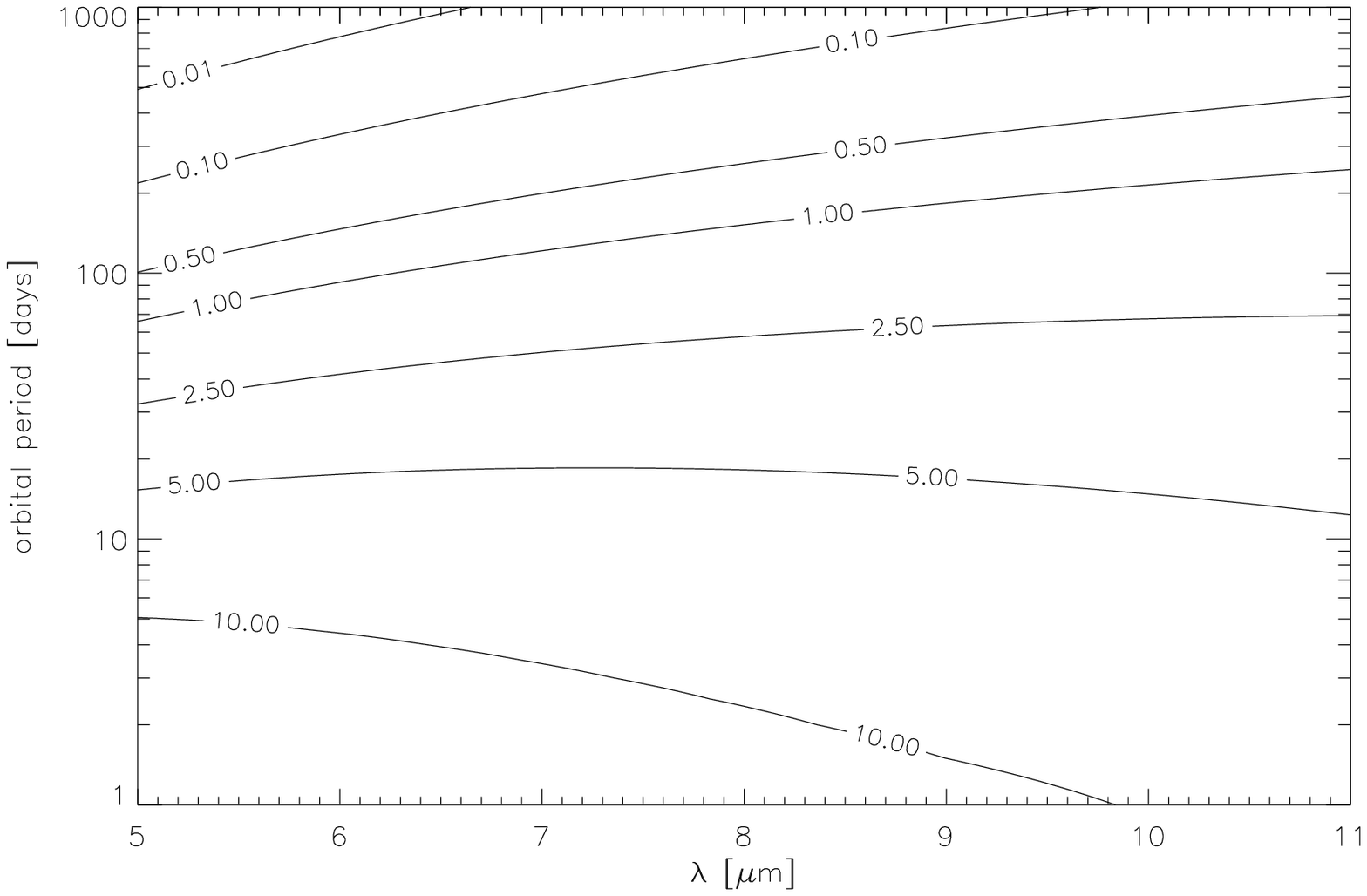} &
    \includegraphics[width=0.49\textwidth]{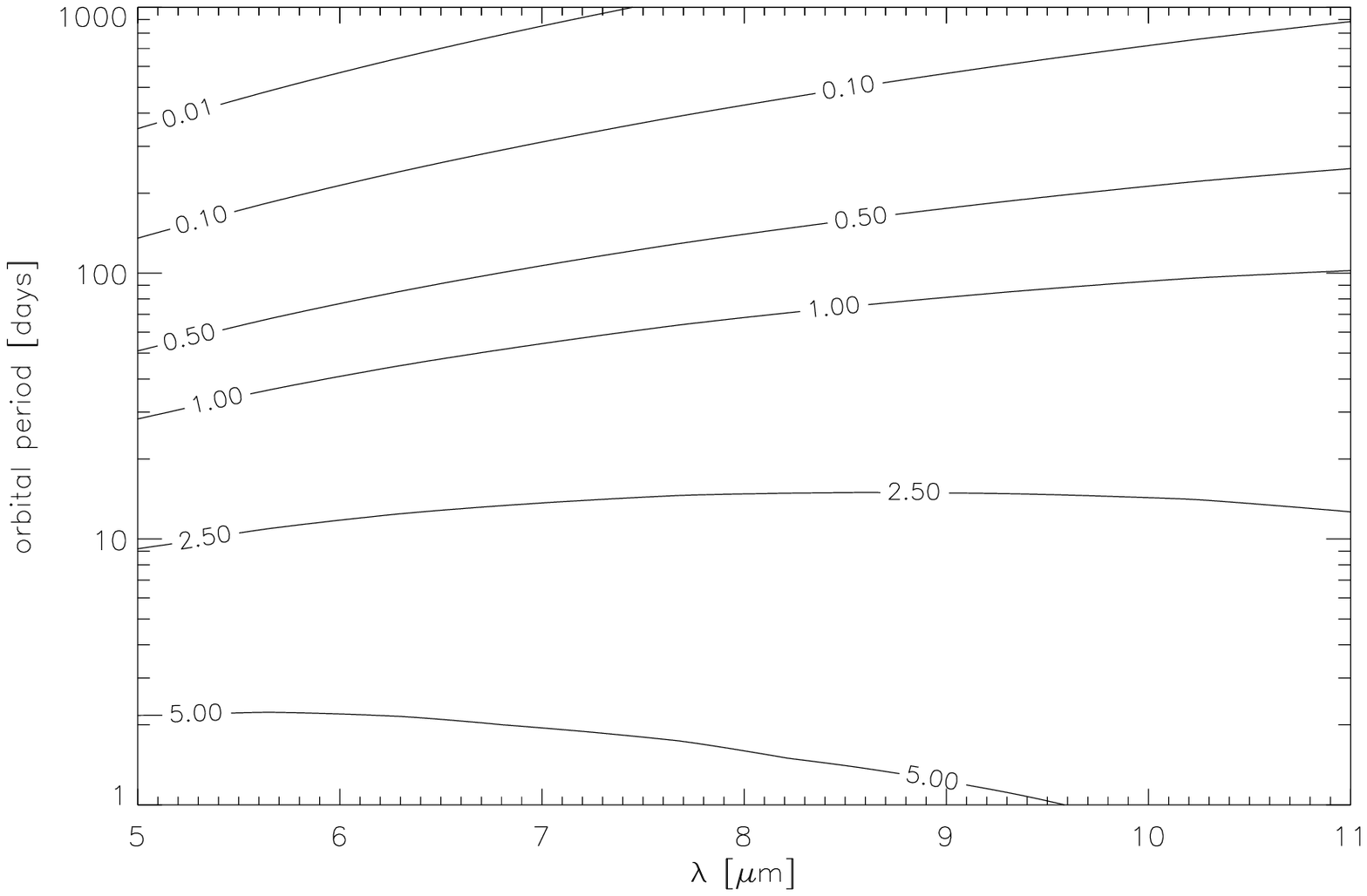} \\

    \includegraphics[width=0.49\textwidth]{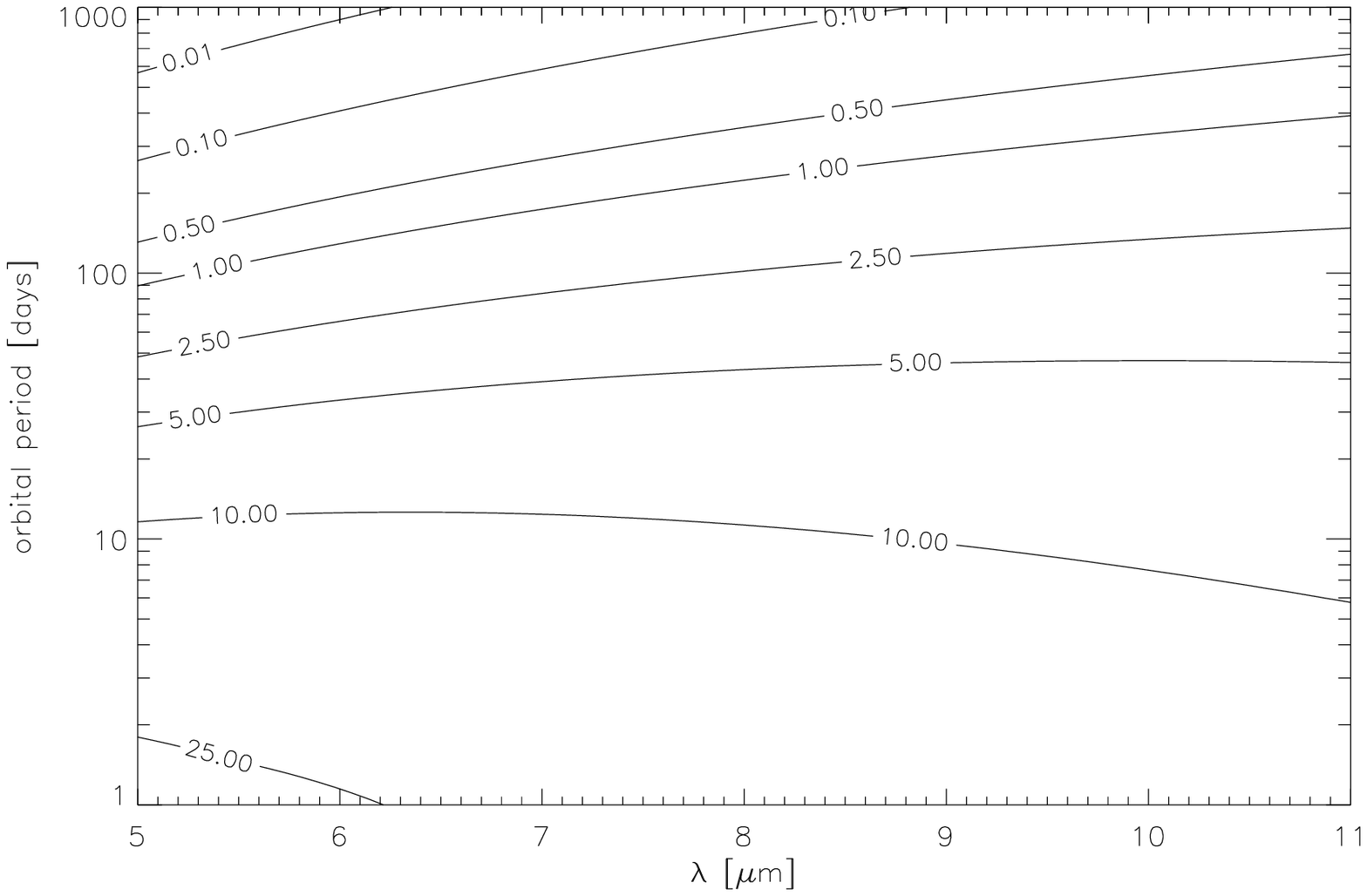} &
    \includegraphics[width=0.49\textwidth]{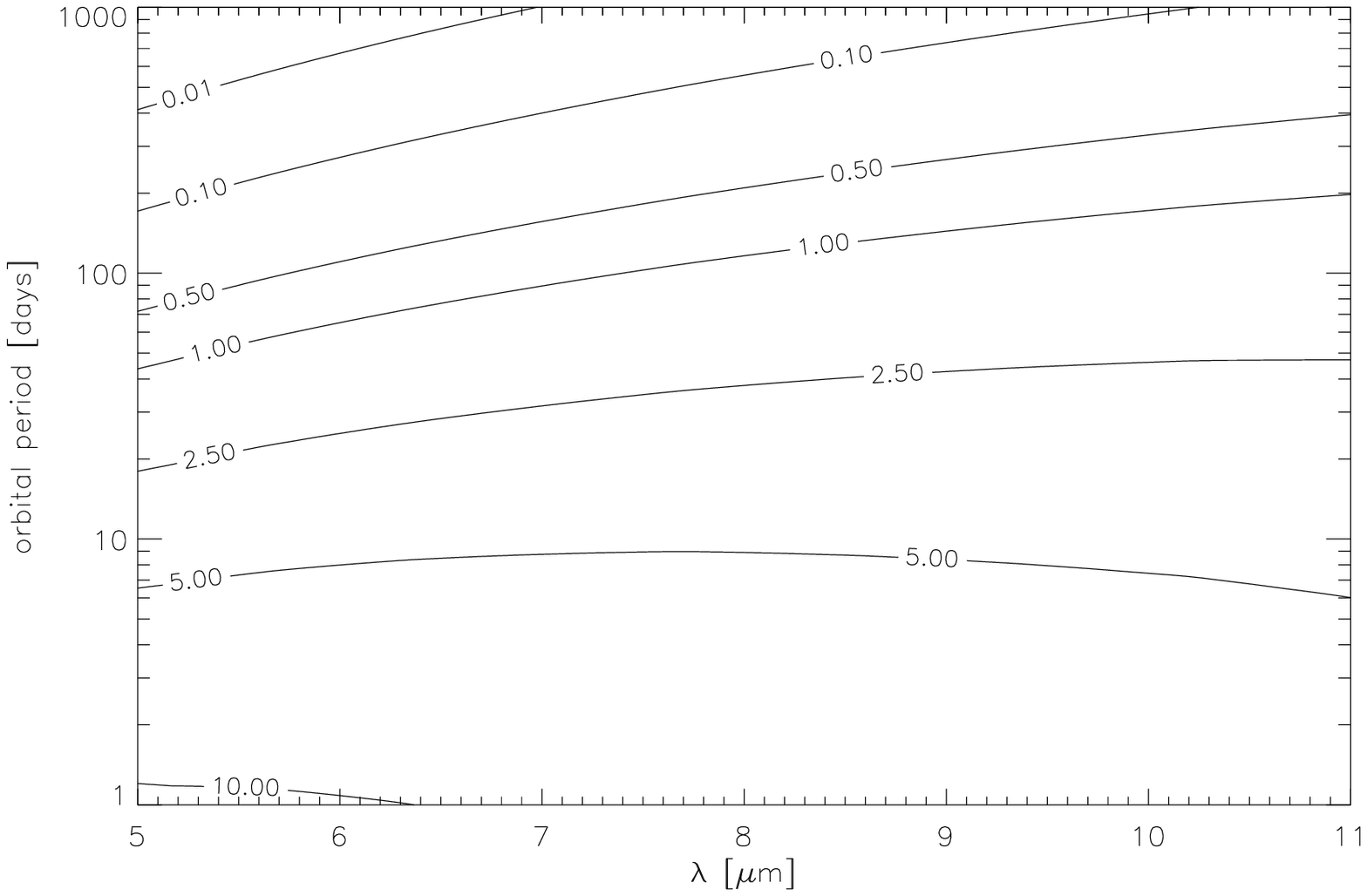}
  \end{array}$
  \caption{\label{f:sec-nep-jup-MIR}S/N for a \emph{single} secondary transit of a Neptune-size planet at 10\,pc (\textbf{top row}) and a Jupiter-size one at 50 pc (\textbf{bottom row}), around a solar-mass star, in the wavelength-planet period space, and at the full resolution of the MIRI low resolution spectrometer ($R\,=\,100$).}
\end{figure*}

As before, for Neptune and Jupiter-mass planets we consider a fiducial spectral feature, with the same $\alpha=0.2$ depth. Results are shown in Figure~\ref{f:sec-nep-jup-MIR}. For jupiters for instance, the S/N is not so sensitive with wavelength as in the primary transit case (Figure~\ref{f:primary-jup-miri-Msun}).

\subsection{Secondary transit - reflection}

\begin{figure*}
  \centering
  \includegraphics[width=0.49\textwidth]{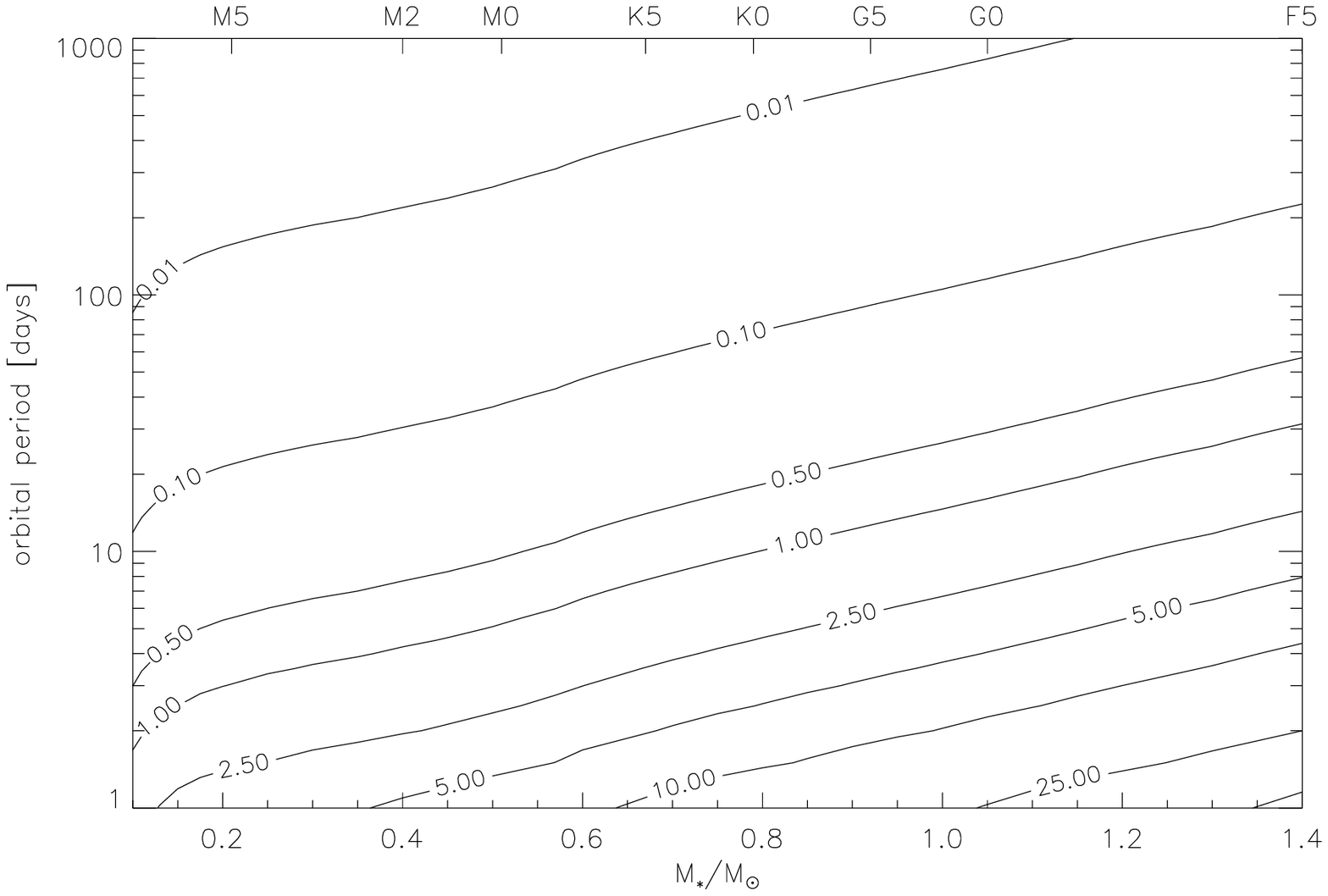}
  \includegraphics[width=0.49\textwidth]{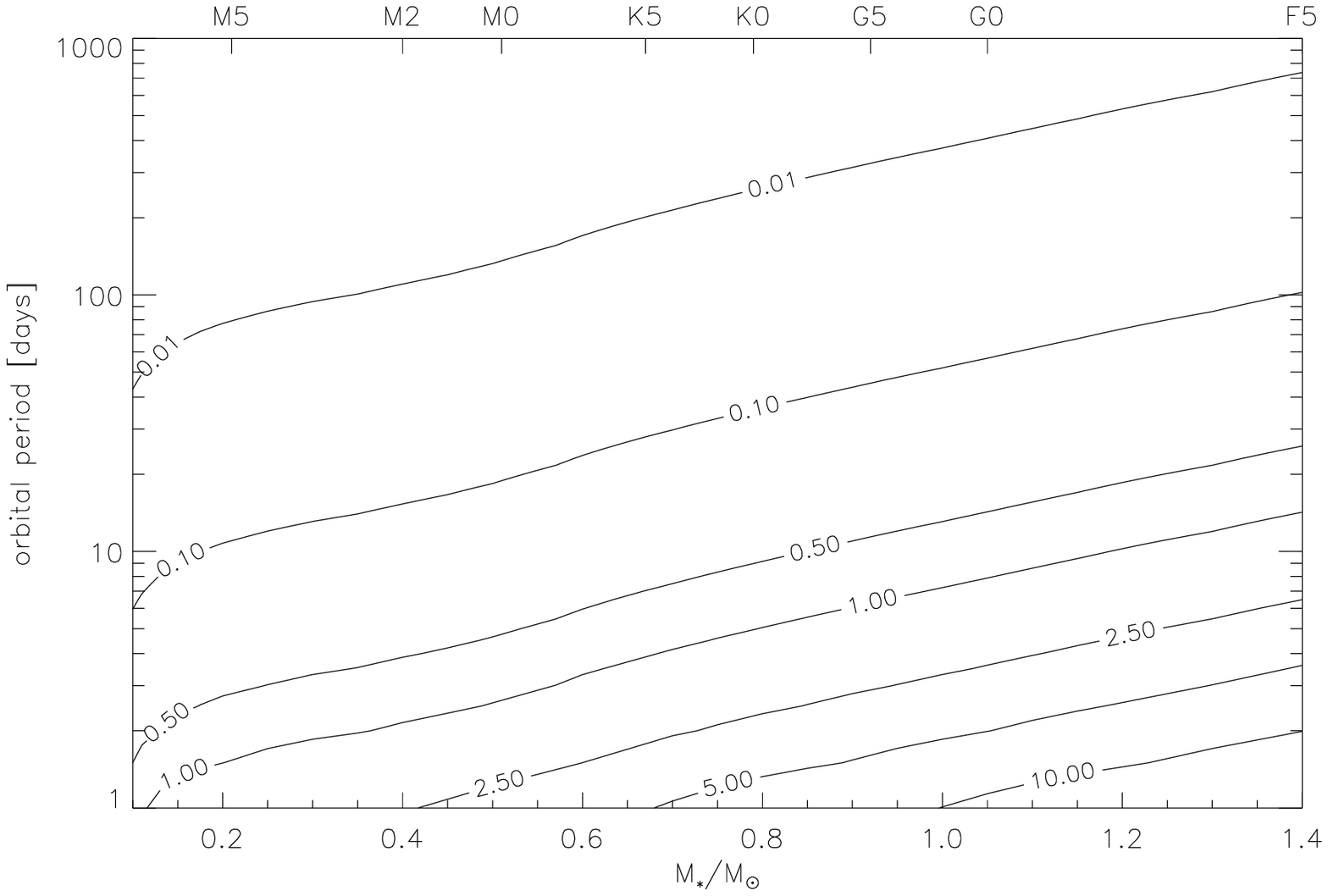}

  \caption{\label{f:reflection-jup}S/N for the detection of a 50\% albedo drop with 0.1\,\um{} effective bandwidth on a Jupiter-mass planet at 50pc with one single transit  at 1\,\um{} (\NIRSpec).}
\end{figure*}

Reflection spectroscopy is considered only with \NIRSpec, because reflection is swamped by thermal emission in the mid-infrared. Reflection spectroscopy of super-earths is not achievable even over the full mission time, and is not shown here. Also, reflection spectroscopy of Jupiter-mass or Neptune-mass planets is achievable in one or a couple of transits only for the most illuminated planets (Figure~\ref{f:reflection-jup} for the Jupiter-mass case).


\section{Test on a \textsl{HST} observation}

In order to test our model, we implemented the corresponding characteristics of the \textsl{Hubble Space Telescope} (HST) \textsl{NICMOS} instrument for observations with the G206 grism (slitless). We simulate the \citet{swain_ch4} observation of HD\,189733\,b. We choose a feature of 2 scale heights (1,400\,K assumed for planetary temperature) and one channel wide ($R=40$) at 2.05\,\um. In this paper, we have assumed that the individual pixel responses are well characterized, enabling to observe very near to the saturation limit. With this setting, our \HST duty cycle is very near 100\%, so we adjust the S/N by the reported 18\% duty cycle of this observation. Our model S/N is 3.6 stronger than the one of the observations, which can account for the pointing oscillations and other optical state variations that we do not model. 

For prediction purposes, we have investigated the S/N for the observation of the primary transit spectroscopy GJ\,1214\,b at the same wavelength and resolution (illustrative of the detection of the CO$_2$ band, 3 scale heights strong). The largest \HST yearly visibility window for GJ\,1214 is 107 days. We consider a planet period of 1.6 days, a transit duration of 48 min, and a \HST orbit period of 96.5 min. We assume that the target is visible only 40\% of each orbit. We compute there are on average 59.6 ($\sigma$=2.6) out of the 68 within-the-window transits (88\%), which are at least partially covered by the visible period of an \HST orbit\footnote{GJ\,1214 has no \textsl{Continuous Viewing Zone (CVZ)} for \HST. Because of atmospheric drag of the low orbit- \HST, it is impossible to know even a couple of weeks beforehand the exact position of the telescope.}. A total of 21.3 ($\sigma$=0.5)  hours of transit are covered by effective orbit portions, and can therefore be cumulated. The required telescope time is at least the double because of the out-of-transit observations for the stellar baseline determination. Applying the above 3.6 scaling factor to the result of our model, we obtain an S/N of 1.8 for $\mu\,=\,18$\,g mol$^{-1}$, and $>5$ for $\mu\,=\,6$. If we consider $\mu\,=\,2$ (consistent with the hydrogen-dominated atmosphere currently assumed to explain the observed radius of the planet), S/N = 2.5 with only one transit. The observation of only one secondary transit with \JWST-\MIRI ($\alpha\,=\,0.2$, $R\,=\,20$ at 10\,\um) would yield a S/N of 3.5. 



\section{Likelihood of occurring targets}
\label{s:stats}

\subsection{Habitable planets}

\begin{figure} 
  \includegraphics[width=\columnwidth]{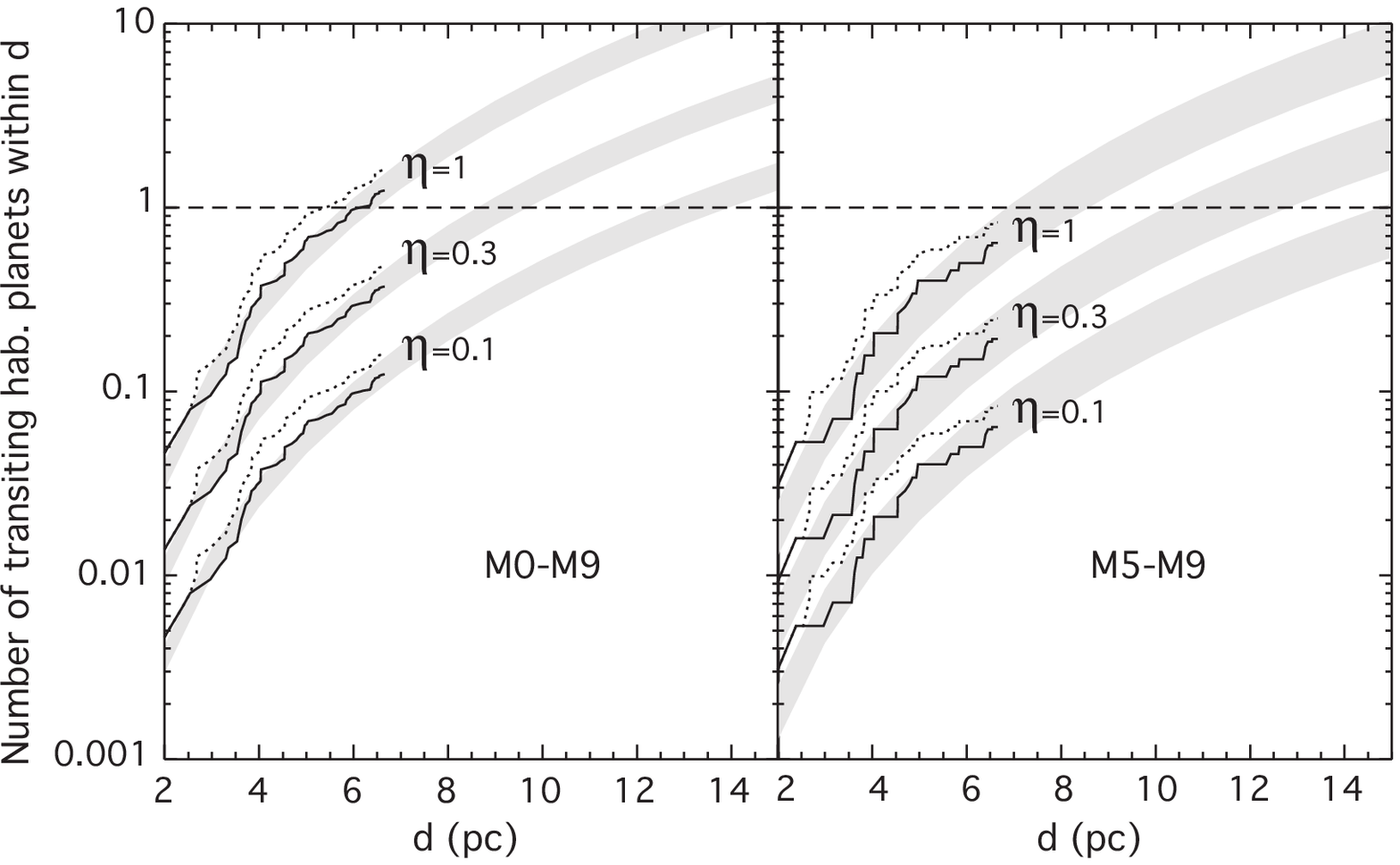} 
  \caption{\label{f:transit-stats}Number of transiting systems receiving Earth's insolation, as a function of distance and mean number of habitable planets per star ($\eta_\oplus$). The doted curves are based on the RECONS (Research Consortium on Nearby Stars) list. The solid curves results from rejecting stars separated from a companion by less then 5\arcsec, as well as well as GJ\,581 (already surveyed for transits) and GL\,876 (eccentric planet in the habitable zone). The gray profiles are a $d^3$ fit ($\pm \sigma$) to the list with close binaries removed.}
\end{figure}

We have seen that transiting habitable super-earths could be characterized at low-resolution with a significant S/N only in the most nearby systems ($<$10\,pc) and when hosted by a M star ($<$0.2\,\Msun) (Figures~\ref{f:4.3-CO2-trans}, \ref{f:O3-primary}, \ref{f:CO2-primary}, \ref{f:emiss-CO2-searth}, \ref{f:emiss-O3} and \ref{f:15um-emission-CO2}). These two criteria match well since the stellar population is dominated by M stars; however, we are dealing with a small number of stars ($\sim$300 within 10~pc). We thus calculate the occurrence likelihood of transiting habitable planets around M stars (Figure~\ref{f:transit-stats}). We make use of a complete census within 6.7\,pc made by RECONS (Research Consortium on Nearby
Stars)\footnote{\href{http://www.chara.gsu.edu/RECONS}{http://www.chara.
gsu.edu/RECONS}}.  For each star, we compute the transit probability for a planet receiving the same insolation as the Earth, and we can thus calculate the statistic number of transiting planets as a function of the distance to the Sun. This value has to be multiplied by $\eta_\oplus$, the mean number of habitable planets per star. Within 10 pc and for M0-M9 stars, $4.6 \times \eta_\oplus$ transiting habitable planets are expected.  We have excluded K, G and F stars because of the low S/N for habitable planets. Including  K stars would increase the statistical number of transits by 10\%. 

One should note also that $\eta_\oplus$ here is not the fraction of stars having a habitable planet because one star may host several habitable planets, which was maybe the case for the Sun 4~Gyrs ago, when Venus, the Earth and Mars were potentially habitable \citep{sels_gl581}. The case of $\eta_\oplus\,>\,1$ is thus not to be discarded.

Because the Earth is located near the inner edge of the habitable zone, using its insolation to compute the transit probability yields optimistic numbers: if we assume a uniform distribution of planets with orbital distance, most habitable planets have a transit probability lower than the one we use. Climate models predict that the outer boundary for the Sun's habitable zone is located around 2\,AU. Using 2\,AU instead of 1\,AU as a reference yields transit probabilities decreased by a factor 2. Therefore, for $\eta_\oplus=0.3$ the statistical number of transiting habitable planets reaches 1 somewhere between 8.5 and 12\,pc for the whole M0-M9 range and between 10 and 16\,pc for the M5-M9 range.

We should also note that our S/N calculations for habitable planets assume a $2R_\oplus$ radius, which may be significantly larger than the average value. For primary transits, the atmospheric signal varies as $R_\RM{p}^{0.8}$ for rocky planets (because density varies with the mass). The S/N scales as $R_\RM{p}^2$ for secondary eclipses. For this reason also, the values we chose seem to represent an optimistic situation where large planets dominate the population of habitable planets.

Last, the computed S/N for a habitable super-Earths assumes an observational
program regularly spread throughout the JWST mission time, and therefore a transiting target known at the beginning of the mission.

\subsection{Jupiters. Neptunes and hot super-earths}

Hot jupiters ($>0.1$ jupiter masses and period $<10$ days) are found by radial  
velocity (RV) surveys around about 2\% of F-G-K stars (see for instance  
\citealt{cumming}). Their period distribution peaks at $\sim 
$3\,days. With these two values we compute the transit probability for  
individual nearby stars of a synthetic  population of nearby stars  
generated with the Besan\c con model \citep{besancon}\footnote{\href{http://model.obs-besancon.fr/}{http://model.obs-besancon.fr}}. We find that the  
statistical number of transiting hot jupiters reaches 1 at 23\,pc. The  
closest known hot jupiter is HD\,189733b, and is found at 20\,pc. We  
should expect about 10 transiting hot jupiter within 50\,pc, among  
which 2 are already known (HD~209458\,b and HAT-P-11\,b). This is why we  
choose this distance for our S/N contours for hot jupiters.

Lower limits on the frequency of longer-period giant planets can be  
estimated from RV \citep{cumming}. A clear trend  
shows that long period jupiters are more frequent than hot ones, with  
at least 10\%  of the stars hosting gas giants with period smaller  
than Jupiter's one. This increase of frequency does not however compensate for the linear decrease of the geometric  
transit probability with the orbital distance. This means that the  
number of transiting cases observed within a certain distance  
decreases as the orbital period increases.

This conclusion may not be true for M stars, for which no hot Jupiter  
have been found but long period gas giant are detected by RV and  
microlensing. Microlensing results  
suggest that $40 \pm 20$\% of the lenses host a massive planet at the  
probed orbital distances (beyond the snow line, \citealt{gould}). The population of lenses is dominated by stars in the 0.3-0.7\,\Msun range. Let $\eta$ be the mean number of planets of considered type per considered star mass range. If we  extrapolate the above result and assume a value 40\% for $\eta$ at the  
snow line\footnote{Taken at 2.7\,AU for the Sun and scaled with luminosity}  
for stars down to 0.1\,\Msun, then the statistical number of transits  
reaches 1 at only 11\,pc. However, despite this potentially high number  
of nearby transiting long-period planets, these systems are hard to  
find with the current methods, and may not be unveiled in time for \JWST.

Short period planets ($<$50\,days) in the 5-20\, Earth mass range have been  
found by the \textsl{HARPS} RV survey around $30 \pm 10$ of the G and K stars  
\citep{mayor_eta_searth}. Taking into account the uncertainty on $\eta$ and  the fact that the dependency of $\eta$ upon the period is yet to be determined, the  statistical number of transiting systems reaches 1 between 8 and 18\,pc  
for G-K stars only, and possibly as close as 5\,pc if extrapolated to M  
dwarfs. These hot low-mass planets  represent promising targets for  
\JWST, as determining if hot super-earths can sustain an atmosphere, and  
of what composition, is a key scientific question.

\section{\New{S/N scaling with various parameters}}

\begin{table*}
  \centering
  \caption{\New{S/N scaling with various parameters}}
  \label{t:scaling}
  \begin{tabular}{lcl}
    \hline\hline
    Parameter                                           &             & Scales as                                              \\
    \hline
    Distance to star                                    & $d$         & $^1/_d$                                                \\
    Collector diameter																	& $D$     		& $D$                                                    \\	
    Number of observed transits (uncorrelated noise)    & $N$         & $\sqrt{N}$                                             \\
    Planet radius  (super-Earths)                       & $R_\RM{p}$  & $R_\RM{p}^{0.8}$ (primary) or $R_\RM{p}^2$ (secondary) \\
    Atmospheric mean molecular mass                                 & $\mu$       & $^1/_\mu$                                              \\
    Number of scale heights (feature opacity)           & $n$         & $n$                                                    \\
    Resolution                                          & $R$         & $^1/_R$                                                \\    
  \hline
  \end{tabular}
\end{table*}

\New{Table \ref{t:scaling} summarizes how the S/N scales with the different parameters considered constant in the contour maps. Each parameter may scale different components of the S/N (for instance, the distance to the star $d$ scales the stellar photon noise S/N).}

\section{Conclusion}

In this paper we have computed S/N for the \emph{detection} of spectral features in exoplanetary atmospheres through eclipse spectroscopy with \JWST. We insist it is a S/N on the detection of spectral feature, and not on the measured value of the flux. The S/N is represented as function of exoplanet parameters (size, insolation, host star, transit duration and frequency) and observational ones (resolution, wavelength). Our spectral features are modeled by a couple of parameters only (strength, width), in order to better explore the parameter space, and identify the regions of interest where detailed atmosphere models and synthetic spectra can bring further insight. We systematically compare a stellar photon noise-only scenario with one containing background and instrumental noises.

One primary transit observation with \JWST-\textsl{NIRSpec}  ($R$=100 mode) will enable to attain S/N = 3 at  3\,\um{} on giant planets. At 50\,pc, this result can be accomplished for jupiters  with periods up to $\sim$300 days around G and F stars (would they be detected), and up to 30 days around K stars. For neptunes  (10\,pc away) the same S/N (same wavelength and resolution) is obtained for planets with periods up to again $\sim$ 300 days but around K and G stars.

In the MIR (\MIRI), at 10\,\um, S/N\,= 3 ($R\,=\,100$) can be obtained by summing $<$10 transits, for neptunes up to 100 days period. For jupiter-mass planets the period limit for MIR S/N\,=\,4 with 4 transits decreases from 30\,days at 5\,\um{} to 4\, days at 11\,\um{} (solar host assumed). In secondary eclipse MIR observations, the limit on the contrary increases with wavelength, from 20 days at 5\,\um{} to 80\,days at 11\,\um{} for jupiter planets. For neptunes, the limit is roughly uniform with wavelength: $\sim$15\,days. In the NIR, neptunes become difficult to characterize even by summing transits, and the jupiter planet limit (S/N\,=\,5 with 1 transit, $R\,=\,100$) is around 4 days and only for wavelengths over 2\,\um. 

In summary, \JWST will better characterize exoplanets already characterized today in eclipse spectroscopy, but, more importantly, will enable to characterize at the level available today (or better) objects more distant, thus more numerous, therefore contributing strongly to comparative planetology.
\\
\\
We devote particular attention to the prospect of characterizing habitable planets. Since achieving a significant S/N requires to sum a large number of transits we systematically compute the cost (fraction of the \JWST mission time) of such observations. Detection of O$_3$ at $\RM{S/N}\,\geq\,3$  is feasible around M4-M9 stars in primary and around M5-M9 in secondary eclipse for warm super-earths 6.7\,pc away or closer, with $\sim$2\% of the 5 year \JWST mission time. We compute that if every M star out to 6.7\,pc where to have one habitable planet, there should be $\sim$1 transiting case. CO$_2$ (15 et 4.3\,\um{} features) are also detectable (only the 4.3 \um{} is not detectable with secondary transits).
Shorter (longer) wavelengths are naturally more suited for primary (secondary) eclipse observations. 

We showed that low mass stars that yield a high S/N on the spectroscopic observation of their habitable planets are also the peak in the number of likely transiting cases. 
Because of the need for radial velocity confirmation, most transit searches survey the brightest stars (magnitude limited samples), while if the purpose is the characterization of atmospheres, we should put more efforts on the most nearby stars, whatever the brightness. We also wish to stress that the census transiting habitable planets around M dwarves should be as complete as possible by the beginning of \JWST operations. Otherwise, we may be faced with the scenario of choosing to start an observational program worth 2\% \footnote{For targets with 100\% yearly visibility} of the 5 year-mission time, only to find after 2.5 years (halfway) that there is a more interesting target.

While a timely precursor to dedicated observatories as \textsl{Darwin} or \textsl{TPF-I/C}, habitable exoplanet eclipse spectroscopy with \JWST can not reach the science objectives of the former. K and G stars are out of reach, and M star planets have a lot of habitability issues \cite{scalo}. Photon collection rate is limited by duration and frequency of the transit, and (not undissociated from the previous) the technique can not give access to enough targets to provide comparative planetology statistics.

Further work will include considering the exact variances of the light curve fitting techniques actually used for the exoplanet spectrum calculation \citep{carter}. We have not modeled limb darkening (for primary transit),  star spots,  and stellar variability which is likely to be peculiar for M dwarves. We recommend detailed preflight characterisation of the detectors. For one, this would enable to mitigate for pointing oscillations. Also, we have assumed here that we work near the saturation limit of the detectors, where their behavior is highly non-linear. Our code is available on request (contact first author).

\begin{acknowledgements}
  We acknowledge support from the European Research Council (Starting Grant 209622: E$_3$ARTH$^\RM{s}$). A.B acknowledges support from Centre National d'Etudes Spatiales. P. Ferruit (CRAL) provided information on NIRSpec, and V. Morreau and S. Ronayette (CEA) provided information on MIRI. The authors also acknowledge discussions with C. Cavarroc, A. Bocaletti, and P.-O. Lagage. \New{We also thank the anonymous referee for comments improving the accuracy of the paper.} We thank A.-S. Maurin for improving the readability.
\end{acknowledgements}

\bibliographystyle{aa}       
	\bibliography{transit-snr} 

\end{document}